\pdfoutput=1
\documentclass[12pt,a4paper]{article}

\usepackage{ifthen} % for conditional statements
\newboolean{pdflatex}
\setboolean{pdflatex}{true} % False for eps figures 

\newboolean{articletitles}
\setboolean{articletitles}{true} % False removes titles in references

\newboolean{uprightparticles}
\setboolean{uprightparticles}{false} %True for upright particle symbols

\newboolean{inbibliography}
\setboolean{inbibliography}{false} %True once you enter the bibliography

\usepackage{microtype}
\usepackage{lineno}  % for line numbering during review
\usepackage{xspace} % To avoid problems with missing or double spaces after
\usepackage{caption} %these three command get the figure and table captions automatically small

\usepackage{graphicx}  % to include figures (can also use other packages)
\usepackage{color}
\usepackage{colortbl}
\usepackage{amsmath} % Adds a large collection of math symbols
\usepackage{amssymb}
\usepackage{amsfonts}
\usepackage{upgreek} % Adds in support for greek letters in roman typeset

\newcommand*\patchAmsMathEnvironmentForLineno[1]{%
\expandafter\let\csname old#1\expandafter\endcsname\csname #1\endcsname
\expandafter\let\csname oldend#1\expandafter\endcsname\csname
end#1\endcsname
 \renewenvironment{#1}%
   {\linenomath\csname old#1\endcsname}%
   {\csname oldend#1\endcsname\endlinenomath}%
}
\newcommand*\patchBothAmsMathEnvironmentsForLineno[1]{%
  \patchAmsMathEnvironmentForLineno{#1}%
  \patchAmsMathEnvironmentForLineno{#1*}%
}
\AtBeginDocument{%
\patchBothAmsMathEnvironmentsForLineno{equation}%
\patchBothAmsMathEnvironmentsForLineno{align}%
\patchBothAmsMathEnvironmentsForLineno{flalign}%
\patchBothAmsMathEnvironmentsForLineno{alignat}%
\patchBothAmsMathEnvironmentsForLineno{gather}%
\patchBothAmsMathEnvironmentsForLineno{multline}%
\patchBothAmsMathEnvironmentsForLineno{eqnarray}%
}

\usepackage{hyperref}    % Hyperlinks in references
\usepackage[all]{hypcap} % Internal hyperlinks to floats.

\usepackage{xspace} 
\usepackage{upgreek}

\def\lhcb {\mbox{LHCb}\xspace}

\def\lhc    {\mbox{LHC}\xspace}

\def\MagUp {\mbox{\em Mag\kern -0.05em Up}\xspace}

\ifthenelse{\boolean{uprightparticles}}%
{

 \def\Pmu         {\ensuremath{\upmu}\xspace}

 \def\Ppsi        {\ensuremath{\uppsi}\xspace}

 \def\PDelta      {\ensuremath{\Delta}\xspace}                 
 \def\PXi      {\ensuremath{\Xi}\xspace}                 
 \def\PLambda      {\ensuremath{\Lambda}\xspace}                 
 \def\PSigma      {\ensuremath{\Sigma}\xspace}                 
 \def\POmega      {\ensuremath{\Omega}\xspace}                 
 \def\PUpsilon      {\ensuremath{\Upsilon}\xspace}                 
 
 \def\PB      {\ensuremath{\mathrm{B}}\xspace}                 
                  
 \def\PD      {\ensuremath{\mathrm{D}}\xspace}

 \def\PJ      {\ensuremath{\mathrm{J}}\xspace}                 
 \def\PK      {\ensuremath{\mathrm{K}}\xspace}

 \def\Pb      {\ensuremath{\mathrm{b}}\xspace}                 
 \def\Pc      {\ensuremath{\mathrm{c}}\xspace}                 
                  
 \def\Pe      {\ensuremath{\mathrm{e}}\xspace}

 \def\Pi      {\ensuremath{\mathrm{i}}\xspace}

}
{

 \def\Pmu         {\ensuremath{\mu}\xspace}

 \def\Ppsi        {\ensuremath{\psi}\xspace}                 
                  
 \mathchardef\PDelta="7101
 \mathchardef\PXi="7104
 \mathchardef\PLambda="7103
 \mathchardef\PSigma="7106
 \mathchardef\POmega="710A
 \mathchardef\PUpsilon="7107
                  
 \def\PB      {\ensuremath{B}\xspace}                 
                  
 \def\PD      {\ensuremath{D}\xspace}

 \def\PJ      {\ensuremath{J}\xspace}                 
 \def\PK      {\ensuremath{K}\xspace}

 \def\Pb      {\ensuremath{b}\xspace}                 
 \def\Pc      {\ensuremath{c}\xspace}                 
                  
 \def\Pe      {\ensuremath{e}\xspace}

 \def\Pi      {\ensuremath{i}\xspace}

}

\makeatletter
\ifcase \@ptsize \relax% 10pt
  \newcommand{\miniscule}{\@setfontsize\miniscule{4}{5}}% \tiny: 5/6
\or% 11pt
  \newcommand{\miniscule}{\@setfontsize\miniscule{5}{6}}% \tiny: 6/7
\or% 12pt
  \newcommand{\miniscule}{\@setfontsize\miniscule{5}{6}}% \tiny: 6/7
\fi
\makeatother

\DeclareRobustCommand{\optbar}[1]{\shortstack{{\miniscule (\rule[.5ex]{1.25em}{.18mm})}
  \\ [-.7ex] $#1$}}

\def\epem       {{\ensuremath{\Pe^+\Pe^-}}\xspace}
\def\mumu       {{\ensuremath{\Pmu^+\Pmu^-}}\xspace}

\def\cquark    {{\ensuremath{\Pc}}\xspace}
\def\cquarkbar {{\ensuremath{\overline \cquark}}\xspace}
\def\ccbar     {{\ensuremath{\cquark\cquarkbar}}\xspace}
\def\bquark    {{\ensuremath{\Pb}}\xspace}
\def\bquarkbar {{\ensuremath{\overline \bquark}}\xspace}
\def\bbbar     {{\ensuremath{\bquark\bquarkbar}}\xspace}

  \def\Kbar    {{\kern 0.2em\overline{\kern -0.2em \PK}{}}\xspace}

\def\KorKbar    {\kern 0.18em\optbar{\kern -0.18em K}{}\xspace}

  \def\Dbar    {{\kern 0.2em\overline{\kern -0.2em \PD}{}}\xspace}

\def\DorDbar    {\kern 0.18em\optbar{\kern -0.18em D}{}\xspace}

\def\Bbar    {{\ensuremath{\kern 0.18em\overline{\kern -0.18em \PB}{}}}\xspace}

\def\BorBbar    {\kern 0.18em\optbar{\kern -0.18em B}{}\xspace}

\def\jpsi     {{\ensuremath{{\PJ\mskip -3mu/\mskip -2mu\Ppsi\mskip 2mu}}}\xspace}
\def\psitwos  {{\ensuremath{\Ppsi{(2S)}}}\xspace}

  \def\Y#1S{\ensuremath{\PUpsilon{(#1S)}}\xspace}% no space before {...}!

\def\Lbar        {{\ensuremath{\kern 0.1em\overline{\kern -0.1em\PLambda}}}\xspace}
\def\LorLbar    {\kern 0.18em\optbar{\kern -0.18em \PLambda}{}\xspace}

\def\BF         {{\ensuremath{\mathcal{B}}}\xspace}

\def\to                 {\ensuremath{\rightarrow}\xspace}

\def\eps   {{\ensuremath{\varepsilon}}\xspace}

\def\AT#1     {\ensuremath{A_{\mathrm{T}}^{#1}}\xspace}           % 2

\def\C#1      {\ensuremath{\mathcal{C}_{#1}}\xspace}                       % 9
\def\Cp#1     {\ensuremath{\mathcal{C}_{#1}^{'}}\xspace}                    % 7
\def\Ceff#1   {\ensuremath{\mathcal{C}_{#1}^{\mathrm{(eff)}}}\xspace}        % 9  
\def\Cpeff#1  {\ensuremath{\mathcal{C}_{#1}^{'\mathrm{(eff)}}}\xspace}       % 7
\def\Ope#1    {\ensuremath{\mathcal{O}_{#1}}\xspace}                       % 2
\def\Opep#1   {\ensuremath{\mathcal{O}_{#1}^{'}}\xspace}                    % 7

\newcommand{\tev}{\ifthenelse{\boolean{inbibliography}}{\ensuremath{~T\kern -0.05em eV}\xspace}{\ensuremath{\mathrm{\,Te\kern -0.1em V}}}\xspace}
\newcommand{\gev}{\ensuremath{\mathrm{\,Ge\kern -0.1em V}}\xspace}
\newcommand{\mev}{\ensuremath{\mathrm{\,Me\kern -0.1em V}}\xspace}
\newcommand{\kev}{\ensuremath{\mathrm{\,ke\kern -0.1em V}}\xspace}
\newcommand{\ev}{\ensuremath{\mathrm{\,e\kern -0.1em V}}\xspace}
\newcommand{\gevc}{\ensuremath{{\mathrm{\,Ge\kern -0.1em V\!/}c}}\xspace}
\newcommand{\mevc}{\ensuremath{{\mathrm{\,Me\kern -0.1em V\!/}c}}\xspace}
\newcommand{\gevcc}{\ensuremath{{\mathrm{\,Ge\kern -0.1em V\!/}c^2}}\xspace}
\newcommand{\gevgevcccc}{\ensuremath{{\mathrm{\,Ge\kern -0.1em V^2\!/}c^4}}\xspace}
\newcommand{\mevcc}{\ensuremath{{\mathrm{\,Me\kern -0.1em V\!/}c^2}}\xspace}

\def\cm   {\ensuremath{\mathrm{ \,cm}}\xspace}

\def\mum  {\ensuremath{{\,\upmu\mathrm{m}}}\xspace}

\def\invnb {\ensuremath{\mbox{\,nb}^{-1}}\xspace}

\def\sec  {\ensuremath{\mathrm{{\,s}}}\xspace}

\def\deriv {\ensuremath{\mathrm{d}}}

\def\gsim{{~\raise.15em\hbox{$>$}\kern-.85em
          \lower.35em\hbox{$\sim$}~}\xspace}
\def\lsim{{~\raise.15em\hbox{$<$}\kern-.85em
          \lower.35em\hbox{$\sim$}~}\xspace}

\def\sPlot{\mbox{\em sPlot}\xspace}
\def\ptot       {\mbox{$p$}\xspace}
\def\pt         {\mbox{$p_{\mathrm{ T}}$}\xspace}

\newcommand{\lum} {\ensuremath{\mathcal{L}}\xspace}
\def\evtgen     {\mbox{\textsc{EvtGen}}\xspace}

\def\geant      {\mbox{\textsc{Geant4}}\xspace}

\def\photos     {\mbox{\textsc{Photos}}\xspace}

\def\pythia     {\mbox{\textsc{Pythia}}\xspace}

\def\tell1  {TELL1\xspace}
\def\ukl1   {UKL1\xspace}

\newcommand{\ie}{\mbox{\itshape i.e.}\xspace}

\usepackage{cite} % Allows for ranges in citations
\usepackage{mciteplus}

\def\mymbarn{\ensuremath{{\mathrm{mb}}}\xspace}
\newcommand{\gevtwofm}{\ensuremath{{\mathrm{\,Ge\kern -0.1em V}^2\!/\mathrm{fm}}}\xspace}
\def\mymub{\ensuremath{{\mathrm{\upmu b}}}\xspace}

\def\Fwd{\ensuremath{\rm Fwd}\xspace}
\def\Bwd{\ensuremath{\rm Bwd}\xspace}
\def\tz{\ensuremath{t_z}\xspace}
\def\Mmumu{\ensuremath{M_{\mu\mu}}\xspace}
\def\corr{\ensuremath{\rm corr}\xspace}
\def\uncorr{\ensuremath{\rm uncorr}\xspace}

\newcommand{\xx}{\ensuremath{\kern 0.5em }}
\def\y {\ensuremath{y}\xspace}
\def\dy {\ensuremath{\deriv\y}\xspace}
\def\dpt {\ensuremath{\deriv\pt}\xspace}

\newcommand{\psimumu}{\ensuremath{\psitwos\to\mumu}\xspace}
\def\pp {\ensuremath{pp}\xspace}
\def\pPb {\ensuremath{p\mathrm{Pb}}\xspace}
\def\pA {\ensuremath{p\mathrm{A}}\xspace}
\def\dAu {\ensuremath{d\mathrm{Au}}\xspace}

\def\sNN {\ensuremath{s_{\mbox{\tiny{\it NN}}}}\xspace}
\def\sNNtitle {\ensuremath{s_{\mbox{\small{\it NN}}}}\xspace}
\def\sqrtsNN {\ensuremath{\sqrt{\sNN}}\xspace}
\def\RpPb{\ensuremath{R_{p\mathrm{Pb}}}\xspace}
\def\RFB{\ensuremath{R_{\mbox{\tiny{FB}}}}\xspace}

\def\sPlot{\mbox{\em sPlot}\xspace}

\usepackage{multirow} % for complicated table
\usepackage{booktabs} % for complicated table
\usepackage{rotating}

\begin{document}

%%%%%%%%%%%%%%%%%%%%%%%%%
%%%%% Title     %%%%%%%%%
%%%%%%%%%%%%%%%%%%%%%%%%%
\renewcommand{\thefootnote}{\fnsymbol{footnote}}
\setcounter{footnote}{1}

% %%%%%%% CHOOSE TITLE PAGE--------
%\onecolumn
%%%%%%%%%%%%%%%%%%%%%%%%%
%%%%%  TITLE PAGE  %%%%%%
%%%%%%%%%%%%%%%%%%%%%%%%%
\begin{titlepage}
\pagenumbering{roman}

% Header ---------------------------------------------------
\vspace*{-1.5cm}
\centerline{\large EUROPEAN ORGANIZATION FOR NUCLEAR RESEARCH (CERN)}
\vspace*{1.5cm}
\noindent
\begin{tabular*}{\linewidth}{lc@{\extracolsep{\fill}}r@{\extracolsep{0pt}}}
\ifthenelse{\boolean{pdflatex}}% Logo format choice
{\vspace*{-2.7cm}\mbox{\!\!\!\includegraphics[width=.14\textwidth]{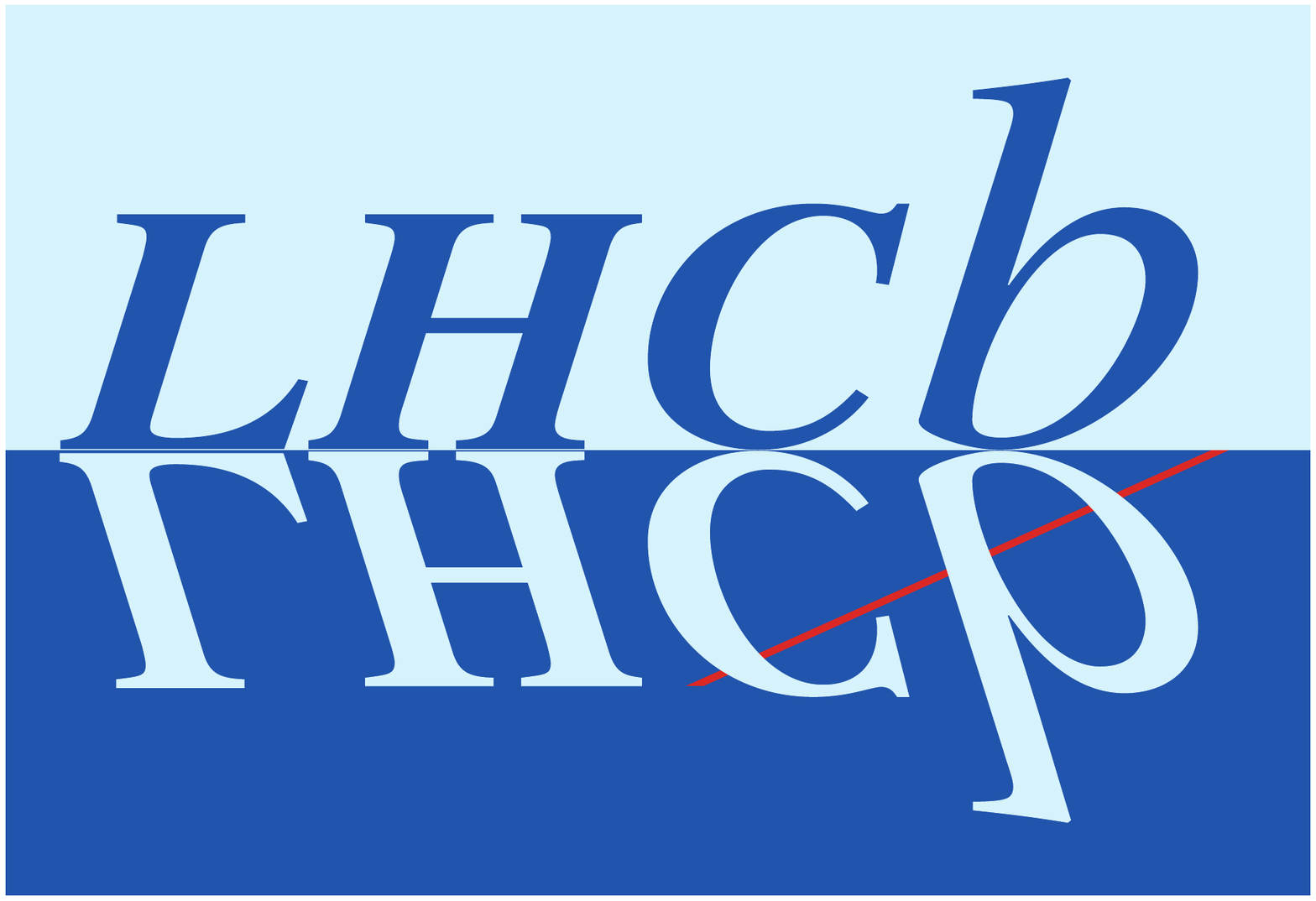}} & &}%
{\vspace*{-1.2cm}\mbox{\!\!\!\includegraphics[width=.12\textwidth]{lhcb-logo.eps}} & &}%
\\
 & & CERN-EP-2016-011 \\  % ID 
 & & LHCb-PAPER-2015-058 \\  % ID 
 & & 29 January 2016\\ % Date - Can also hardwire e.g.: 23 March 2010
\end{tabular*}

\vspace*{2.0cm}

% Title --------------------------------------------------
{\normalfont\bfseries\boldmath\huge
\begin{center}
  Study of $\psitwos$ production and cold nuclear matter effects in $\pPb$ collisions at $\sqrt{\sNNtitle}=5\tev$
\end{center}
}

\vspace*{1.0cm}

% Authors -------------------------------------------------
\begin{center}
The LHCb collaboration\footnote{Authors are listed at the end of this paper.}
\end{center}

\vspace{\fill}

% Abstract -----------------------------------------------
\begin{abstract}
  \noindent
  The production of $\psitwos$ mesons is studied in dimuon final states using proton-lead collision data collected by the LHCb detector. The data sample corresponds to an integrated luminosity of $1.6\invnb$. The nucleon-nucleon centre-of-mass energy of the proton-lead collisions is $\sqrt{\sNN}=5\tev$. The measurement is performed using \psitwos mesons with transverse momentum less than $14\gevc$ and rapidity $y$ in the ranges $1.5<y<4.0$ and $-5.0<y<-2.5$ in the nucleon-nucleon centre-of-mass system. The forward-backward production ratio and the nuclear modification factor are determined for $\psitwos$ mesons. Using the production cross-section results of \psitwos and \jpsi mesons from $b$-hadron decays, the $\bbbar$ cross-section in \pPb collisions at $\sqrt{\sNN}=5\tev$ is obtained.
\end{abstract}

\vspace*{2.0cm}

\begin{center}
  Submitted to JHEP
\end{center}

\vspace{\fill}

{\footnotesize 
\centerline{\copyright~CERN on behalf of the \lhcb collaboration, licence \href{http://creativecommons.org/licenses/by/4.0/}{CC-BY-4.0}.}}
\vspace*{2mm}

\end{titlepage}

%%%%%%%%%%%%%%%%%%%%%%%%%%%%%%%%
%%%%%  EOD OF TITLE PAGE  %%%%%%
%%%%%%%%%%%%%%%%%%%%%%%%%%%%%%%%

%  empty page follows the title page ----
\newpage
\setcounter{page}{2}
\mbox{~}

\cleardoublepage

% %%%%%%%%%%%%% ---------

\renewcommand{\thefootnote}{\arabic{footnote}}
\setcounter{footnote}{0}

%%%%%%%%%%%%%%%%%%%%%%%%%%%%%%%%
%%%%%  Table of Content   %%%%%%
%%%%%%%%%%%%%%%%%%%%%%%%%%%%%%%%
%%%% Uncomment next 2 lines if desired
%\tableofcontents
%\cleardoublepage

%%%%%%%%%%%%%%%%%%%%%%%%%
%%%%% Main text %%%%%%%%%
%%%%%%%%%%%%%%%%%%%%%%%%%

\pagestyle{plain} % restore page numbers for the main text
\setcounter{page}{1}
\pagenumbering{arabic}

% Main body of the paper
\section{Introduction}
\label{sec:Introduction}
\noindent 
 The quark-gluon plasma (QGP) is a state of matter with asymptotically free partons,
which is expected to exist at extremely high temperature and density.
It is predicted that heavy quarkonium production will be significantly suppressed
in ultrarelativistic heavy-ion collisions if a QGP is created~\cite{Matsui:1986dk}. 
This suppression is regarded as one of the most important signatures for the formation of the QGP.
Heavy quarkonium production can also be suppressed in proton-nucleus (\pA) collisions, where hot nuclear matter, \ie QGP, is not expected to be created 
and only cold nuclear matter (CNM) effects exist.
Such CNM effects include: initial-state nuclear effects on the parton densities (shadowing);
coherent energy loss consisting of initial-state parton energy loss and final-state energy loss;
and final-state absorption by nucleons, which is expected to be negligible at LHC energies~\cite{Eskola:2009uj,Ferreiro:2013pua,Arleo:2012rs,Albacete:2013ei,Adeluyi:2013tuu,Chirilli:2012jd,Chirilli:2012sk,Arleo:2013zua}.
The study of \pA collisions is important to disentangle the effects of 
QGP from those of CNM,
and to provide essential input to the understanding of nucleus-nucleus collisions. 

Nuclear effects are usually characterized by the nuclear modification factor,
defined as the production cross-section of a given particle per nucleon in \pA collisions
divided by that in proton-proton ($pp$) collisions, 
\begin{equation}
\label{eq:AttenuationFactor}
 R_{\pA}(y,\pt,\sqrt{s_{\mbox{\tiny{\it NN}}}})
    \equiv\frac{1}{A}
    \frac{\deriv^2\sigma_{\pA}(y,\pt,\sqrt{\sNN})/\dy\dpt}
         {\deriv^2\sigma_{pp}(y,\pt,\sqrt{\sNN})/\dy\dpt},
\end{equation}
where $A$ is the atomic mass number of the nucleus, 
$y$ (\pt) is the rapidity (transverse momentum) of the produced particle, 
and \sqrtsNN is the centre-of-mass energy of the proton-nucleon system.
Throughout this paper, $y$ always indicates the rapidity in the nucleon-nucleon centre-of-mass system.

The suppression of quarkonium and light hadrons at large rapidity has been observed 
in \pA collisions~\cite{Leitch:1999ea,Alessandro:2006jt,Abt:2006va,Abt:2008ya}
and in deuteron-gold collisions~\cite{Arsene:2004ux,Adler:2004eh,Adare:2010fn,Adare:2012bv,Adare:2013ezl}. 
The proton-lead (\pPb) collisions recorded at the LHC in 2013 enable the study of 
CNM effects at the \tev scale.
With these \pPb data, the production cross-sections of prompt \jpsi mesons, \jpsi mesons from $b$-hadron decays, and $\PUpsilon$ mesons were measured, 
and the CNM effects were studied by determining 
the nuclear modification factor \RpPb and the forward-backward production 
ratio \RFB~\cite{LHCb-PAPER-2013-052,LHCb-PAPER-2014-015}.
Working in the nucleon-nucleon rest frame, the ``forward'' and ``backward'' directions are defined with respect to the direction of the proton beam.
The ratio \RFB is defined as 
\begin{equation}
\label{eq:RFB}
 \RFB(y,\pt,\sqrt{\sNN})\equiv
\frac{\sigma_{\pPb}(+|y|,\pt,\sqrt{\sNN})}
     {\sigma_{\pPb}(-|y|,\pt,\sqrt{\sNN})}.
\end{equation}
The advantage of measuring this ratio is that it does not rely on 
knowledge of the production cross-section in $pp$ collisions.
Furthermore, part of the experimental systematic uncertainties 
and theoretical scale uncertainties cancel in the ratio.

Previous measurements in fixed-target \pA collisions by E866~\cite{Leitch:1999ea},
NA50~\cite{Alessandro:2006jt} and HERA-B~\cite{Abt:2006va} showed that
the production cross-sections for both \jpsi and \psitwos mesons 
are suppressed in \pA collisions compared with those in $pp$ collisions. 
These measurements also showed stronger suppression at central rapidity
for \psitwos mesons than for \jpsi mesons,
while at forward rapidity the suppressions were compatible within large uncertainties.
The PHENIX experiment made similar observations in \dAu collisions at RHIC~\cite{Adare:2013ezl}.
The ALICE experiment measured the \psitwos suppression in \pPb collisions at the \lhc~\cite{Abelev:2014zpa}.
Nuclear shadowing and energy loss predict equal suppression of \jpsi and \psitwos mesons,
and so cannot explain the observations.
One explanation is that the charmonium states 
produced at central rapidity spend more time in the medium than those at forward rapidities;
therefore the loosely bound \psitwos mesons are more easily 
suppressed than \jpsi mesons at central rapidity~\cite{Vogt:2001ky,Kopeliovich:1991pu,McGlinchey:2012bp}.
In this picture it is expected that the charmonium states will spend a much shorter time in the CNM 
at LHC energies than at lower energies, 
leading to similar suppression for \psitwos and \jpsi mesons even at central rapidity.

The excellent reconstruction resolution of the \lhcb detector for primary and secondary vertices~\cite{Alves:2008zz} 
provides the ability to separate prompt $\psitwos$ mesons,
which are produced directly from \pp collisions,
from those originating from $b$-hadron decays (called ``$\psitwos$ from $b$" in the following). 
In this analysis, the production cross-sections of prompt \psitwos mesons
and \psitwos from $b$
are measured in \pPb collisions at $\sqrt{\sNN}=5\tev$.
The nuclear modification factor \RpPb and the forward-backward production ratio $\RFB$ are determined in the range $2.5<|y|<4.0$.
Using the production cross-sections of \psitwos from $b$ and \jpsi from $b$, the \bbbar production cross-section in \pPb collisions is obtained.

\section{Detector and datasets}
\label{sec:Detector}
The \lhcb detector~\cite{Alves:2008zz,LHCb-DP-2014-002} is a single-arm forward
spectrometer covering the \mbox{pseudorapidity} range $2<\eta <5$,
designed for the study of particles containing \bquark or \cquark
quarks. 
The detector includes a high-precision tracking system
consisting of a silicon-strip vertex detector surrounding the $\pPb$
interaction region, a large-area silicon-strip detector located
upstream of a dipole magnet with a bending power of about
$4{\mathrm{\,Tm}}$, and three stations of silicon-strip detectors and straw
drift tubes placed downstream of the magnet.
The tracking system provides a measurement of momentum, \ptot, of charged particles with
a relative uncertainty that varies from 0.5\% at low momentum to 1.0\% at 200\gevc.
The minimum distance of a track to a primary vertex, the impact parameter, is measured with a resolution of $(15+29/\pt)\mum$,
where \pt is the component of the momentum transverse to the beam, in\,\gevc.
Different types of charged hadrons are distinguished using information
from two ring-imaging Cherenkov detectors. 
Photons, electrons and hadrons are identified by a calorimeter system consisting of
scintillating-pad and preshower detectors, an electromagnetic
calorimeter and a hadronic calorimeter. Muons are identified by a
system composed of alternating layers of iron and multiwire
proportional chambers.
The online event selection is performed by a trigger, 
which consists of a hardware stage, based on information from the calorimeter and muon
systems, followed by a software stage, which applies a full event
reconstruction.

With the proton beam travelling in the direction from the vertex detector to the muon system and the lead beam circulating in the opposite direction, the LHCb spectrometer covers forward rapidities. With reversed beam directions backward rapidities are accessible.
The data sample used in this analysis is collected from the \pPb collisions in early 2013,
corresponding to an integrated luminosity of $1.1\invnb$ ($0.5\invnb$) for forward (backward) collisions. 
The instantaneous luminosity was around $5\times10^{27}~\cm^{-2}\sec^{-1}$, five orders
of magnitude below the nominal LHCb luminosity for \pp collisions. 
Therefore, the data were taken using a hardware trigger which simply rejected empty events.
The software trigger for this analysis required one well-reconstructed track with hits in the muon system and \pt greater than $600\mevc$.

Simulated samples based on \pp collisions at $8\tev$ are used to determine the acceptance and reconstruction efficiencies. The simulation samples are reweighted so that the track multiplicity distribution reproduces the experimental data of \pPb collisions at $5\tev$.
In the simulation, $pp$ collisions are generated using \pythia~\cite{Sjostrand:2006za}
with a specific \lhcb configuration~\cite{LHCb-PROC-2010-056}.  Decays of hadronic particles are described by \evtgen~\cite{Lange:2001uf}, in which final-state radiation is generated using \photos~\cite{Golonka:2005pn}. The interaction of the generated particles with the detector, and its response, are implemented using the \geant toolkit~\cite{Allison:2006ve, *Agostinelli:2002hh} as described in Ref.~\cite{LHCb-PROC-2011-006}.

\section{Event selection and cross-section determination}
\label{sec:psi2SEventSelectionXsection}
The measurement of $\psitwos$ production is based on the method described in
Refs.~\cite{LHCb-PAPER-2011-003,LHCb-PAPER-2013-016,LHCb-PAPER-2013-052}.
The $\psitwos$ candidates are reconstructed using dimuon final states
from events with at least one primary vertex.
The tracks should be of good quality, have opposite sign charges and be identified as muons with high \pt.
The two muon tracks are required to originate from a common vertex with good vertex fit quality,
and the reconstructed \psitwos mass should be in the range $\pm145\mevcc$ around the known \psitwos mass~\cite{PDG2014}.

Due to the small size of the data sample, 
only one-dimensional differential cross-sections are measured.
The differential production cross-section of $\psitwos$ mesons in a given kinematic bin is defined as 
\begin{equation}
\label{eq:psi2SDifferential}
\frac{\deriv\sigma}{\deriv{}X} =\frac{N}
{\lum\times\BF\times\Delta{}X},
\end{equation}
where $X$ denotes \pt or \y, $N$ is the efficiency-corrected number 
of $\psitwos$ signal candidates reconstructed with the dimuon final state 
in the given bin of $X$, $\Delta{}X$ is the bin width, \lum is the integrated luminosity, 
and $\BF$ is the branching fraction of the $\psimumu$ decay, $\BF(\psimumu)=(7.9\pm0.9)\times10^{-3}$~\cite{PDG2014}.
Assuming lepton universality in electromagnetic decays, this branching fraction is replaced by that of the $\psitwos\to\epem$, which has a much smaller uncertainty,
$\BF(\psitwos\to\epem)=(7.89\pm0.17)\times10^{-3}$~\cite{PDG2014}.

The integrated luminosity of the data sample used in this analysis was determined using a van der Meer scan, and calibrated separately for the \pPb forward and backward samples~\cite{LHCb-PAPER-2014-047}.
The kinematic region of the measurement is 
$\pt<14\gevc$ and $1.5<y<4.0$ ($-5.0<y<-2.5$) for the forward (backward) sample.
For the single differential cross-section measurements, the transverse
momentum range $\pt<14~\gevc$ is divided into five bins with edges at
$(0,~2,~3,~5,~7,~14)\gevc$.
The rapidity range is divided into five bins of width $\Delta\y=0.5$.

%%%%%%%%%%%%%%%%%%%%%%%%%%%%%%%%%%%%%%
\section{Signal extraction and efficiencies}
\label{sec:SignalExtraction}
The numbers of prompt $\psitwos$ and \psitwos from \bquark in each kinematic bin are determined from an extended unbinned maximum likelihood fit
performed simultaneously to the distributions of the dimuon invariant mass \Mmumu and the pseudo proper decay time \tz~\cite{LHCb-PAPER-2011-003},
defined as
\begin{equation}
\label{psiPseudoProperTime}
t_z=\frac{(z_\psi-z_{\mathrm{PV}})\times{}M_\psi}{p_z},
\end{equation}
where $z_\psi$\/ is the position of the \psitwos\ decay vertex along the beam axis,
$z_{\mathrm{PV}}$\/ that of the primary vertex refitted after removing
the two muon tracks from the \psitwos\ candidate, $p_z$\/ the 
$z$ component of the measured \psitwos\ momentum,
and $M_\psi$\/ the known \psitwos\ mass~\cite{PDG2014}.

The invariant mass distribution of the signal in each bin is modelled by a
Crystal Ball (CB) function~\cite{Skwarnicki:1986xj}, 
where the tail parameters are fixed to the values found in simulation and the other parameters are allowed to vary.
For differential cross-section measurements, the sample size in each bin is very small. Therefore, in order to stabilise the fit, the mass resolution of the CB function is fixed to the value obtained from the \jpsi sample, scaled by the ratio of the \psitwos mass to the \jpsi mass.
The invariant mass distribution of the combinatorial background is described by an exponential function with variable slope parameter.
The signal distribution of $t_z$ can be described~\cite{CERN-THESIS-2010-307} by a $\delta$-function at $t_z=0$
for prompt \psitwos and an exponential function for the component of \psitwos from $b$,
both convolved with a Gaussian resolution function.
The width of the resolution function and the slope of the exponential function are free in the fit.
The background distribution of $t_z$ in each kinematic bin is modelled with an empirical function determined from sidebands of the invariant mass distribution.

\begin{figure}[!htb]
\begin{center}
\includegraphics[width=0.42 \textwidth]{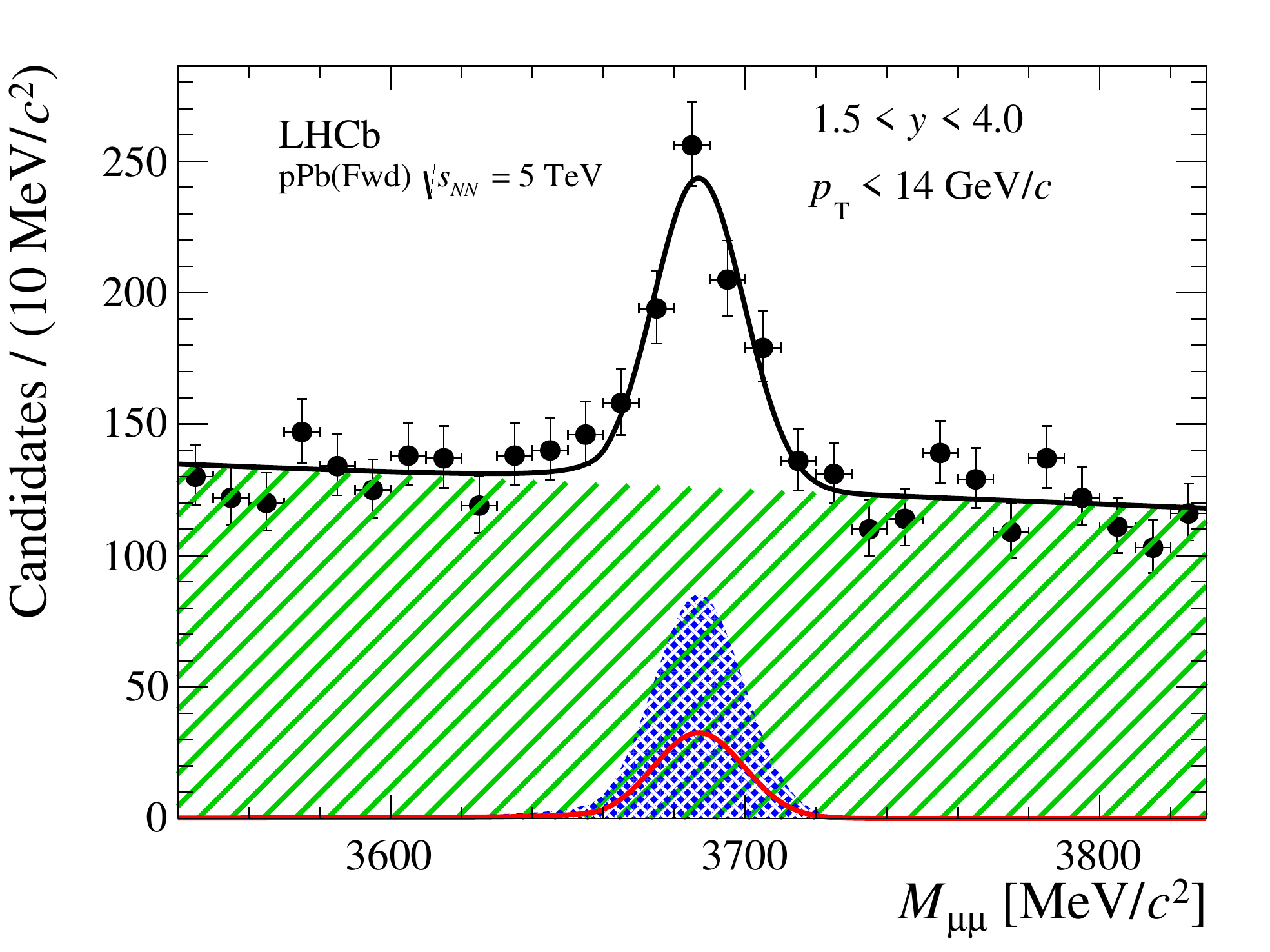}
\includegraphics[width=0.42 \textwidth]{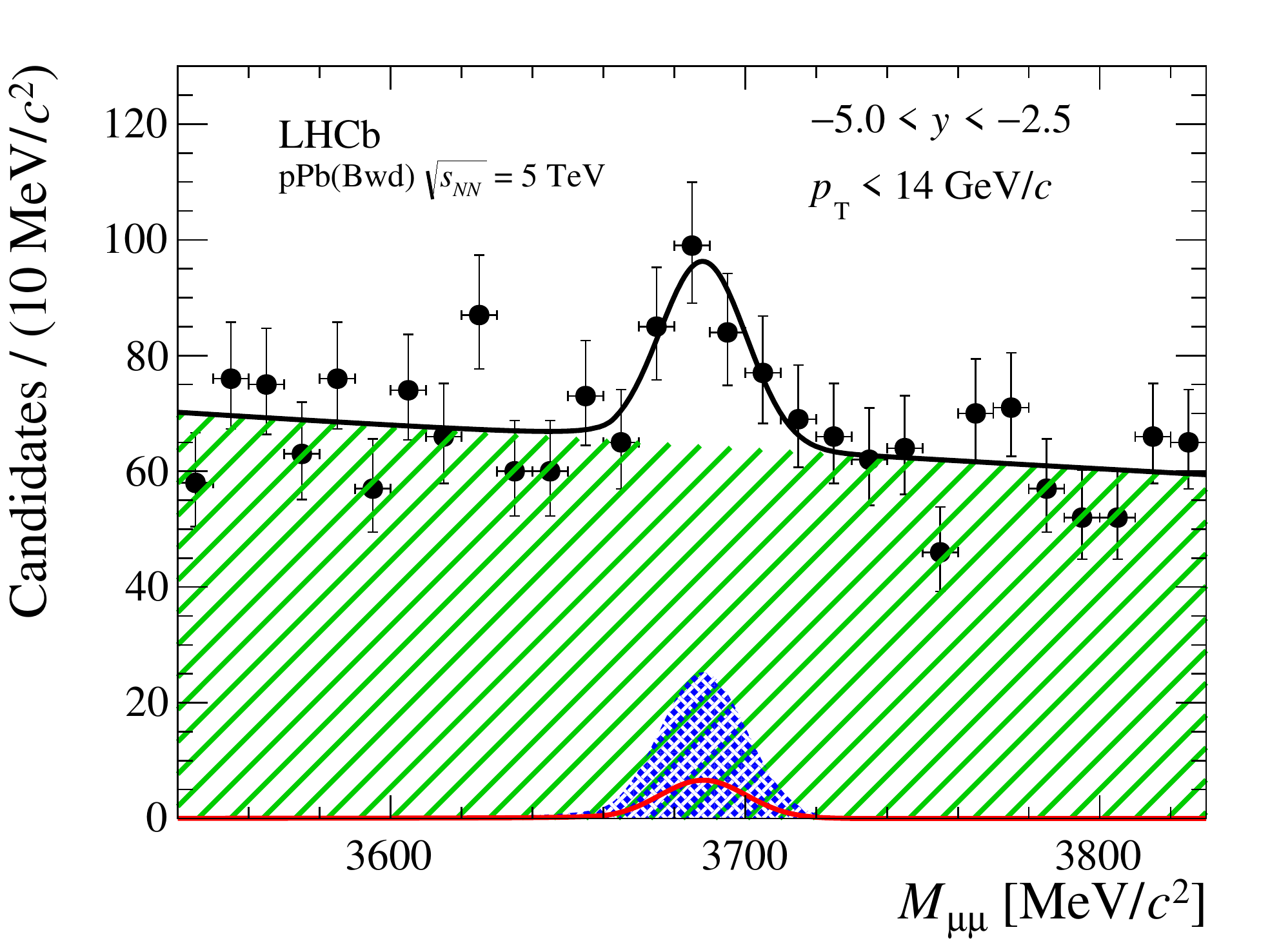}
\includegraphics[width=0.42 \textwidth]{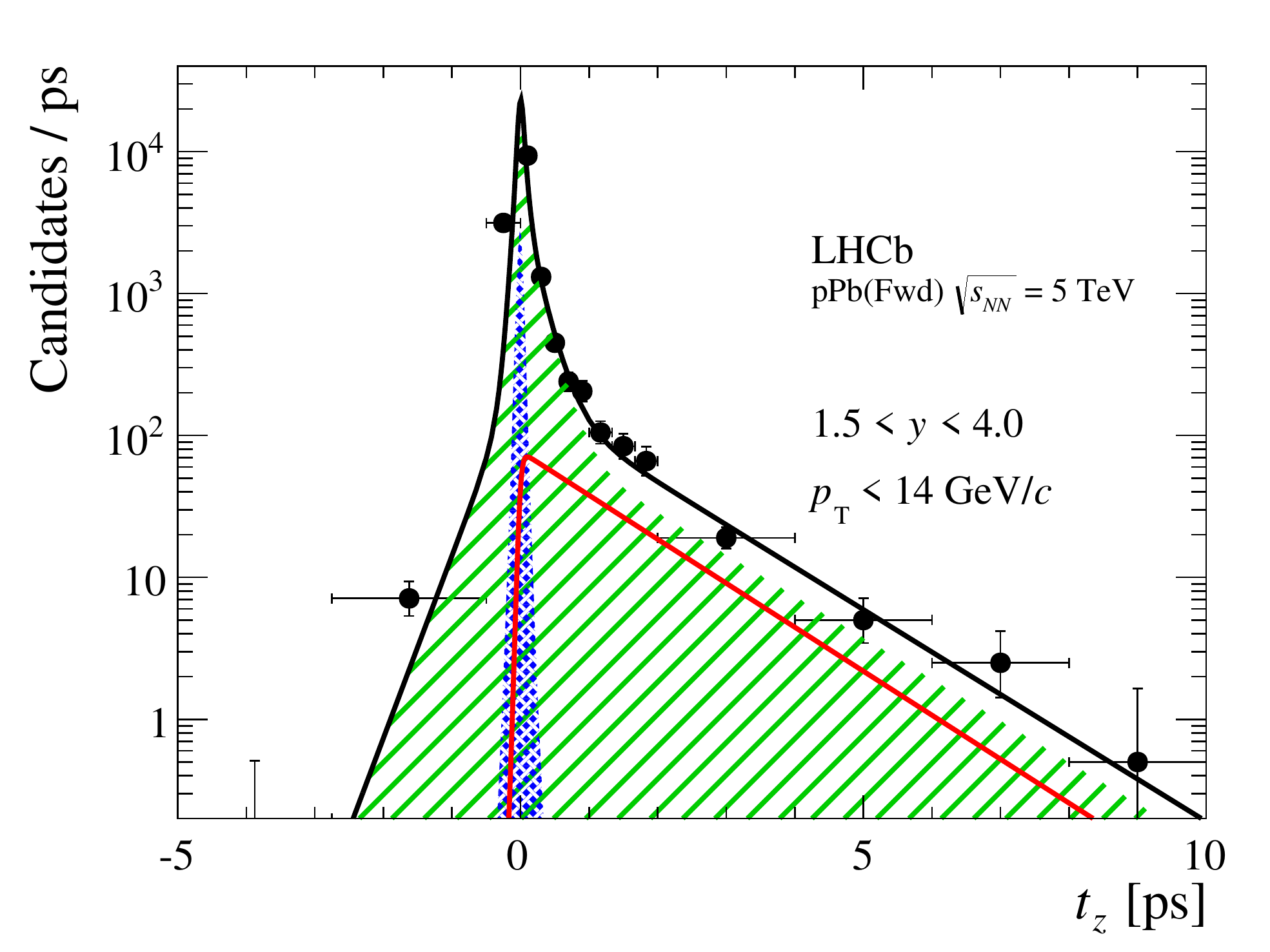}
\includegraphics[width=0.42 \textwidth]{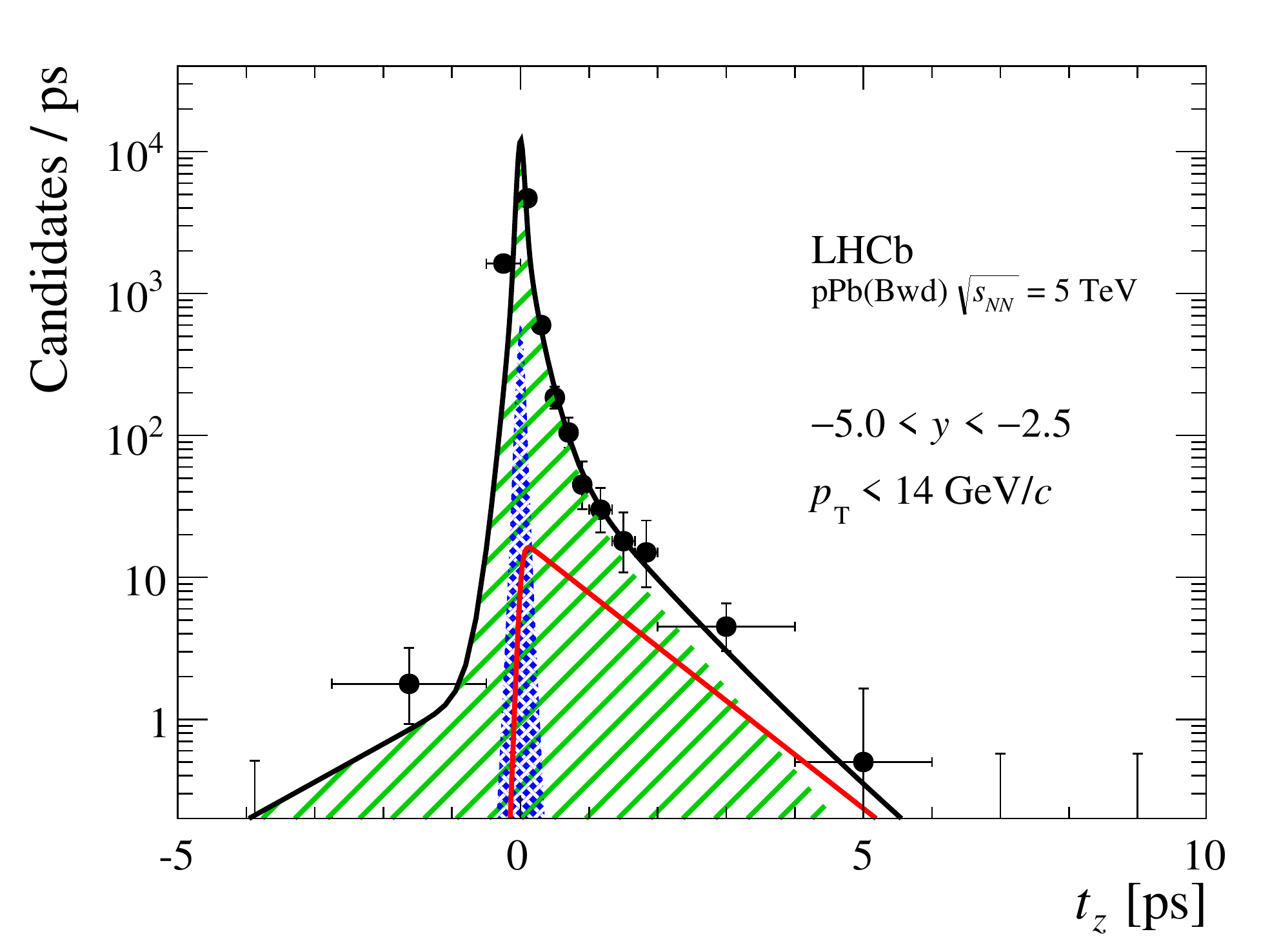}
\vspace*{-0.5cm}
\end{center}
\caption{\small 
    Projections of the fit results to (top) the dimuon invariant mass \Mmumu
    and (bottom) the pseudo proper decay time \tz
    in (left) $\pPb$ forward and (right) backward data. In all plots
    the total fitted function is shown by the (black) solid line,
    the combinatorial background component is shown as the (green) hatched area,
    the prompt signal component by the (blue) shaded area, 
    and the $b$-component by the (red) light solid line.
    }
\label{fig:tzFit}
\end{figure}

Figure \ref{fig:tzFit} shows projections of the fit to \Mmumu and \tz 
for the full $\pPb$ forward and backward samples.
The combinatorial background in the backward region is higher than that in the forward region, because the track multiplicity in the backward region is larger.
The mass resolution is $13\mevcc$ for both the forward and backward samples.
The total estimated signal yield for prompt \psitwos mesons in the forward (backward) sample
is $285\pm34$ ($81\pm23$), and that for \psitwos from \bquark in the forward (backward) sample is $108\pm16$ ($21\pm8$), where the uncertainties are statistical only.

The efficiency-corrected signal yield $N$ is obtained from 
the sum of $w_i/\eps_i$ over all candidates in the given bin.
The weight $w_i$ is obtained with the \sPlot technique using \Mmumu and \tz 
as discriminating variables~\cite{Pivk:2004ty}.
The total efficiency $\eps_i$, which depends on \pt and \y, includes the geometrical acceptance,
the reconstruction efficiency, the muon identification efficiency,
and the trigger efficiency.
The acceptance and reconstruction efficiencies are determined from simulation, assuming that the produced \psitwos mesons are unpolarised.
The efficiency of the muon identification and the trigger efficiency are obtained from data using a tag-and-probe method as described below.

%%%%%%%%%%%%%%%%%%%%%%%%%%%%%%%%%%%%%%%%%%%%%%%%%%%%%%%
\section{Systematic uncertainties}
\label{sec:systematics}
Several sources of systematic uncertainties affecting 
the production cross-section measurements are discussed in the following
and summarised in Table \ref{tab:psi2SSystematics}.

The uncertainty on the muon track reconstruction efficiency is studied with a data-driven tag-and-probe method, using a \jpsi sample in which one muon track is fully reconstructed while the other one is reconstructed using only specific sub-detectors~\cite{LHCb-DP-2013-002}.
Taking into account the difference of the track multiplicity distribution between data and simulation, the total uncertainty is found to be $1.5\%$. 

The uncertainty due to the muon identification efficiency 
is assigned to be $1.3\%$ for both the forward and backward samples as obtained in the \jpsi analysis in \pPb collisions~\cite{LHCb-PAPER-2013-052}.
It is estimated using \jpsi candidates reconstructed with one muon identified by the muon system and the other identified by selecting a track depositing the energy of a minimum-ionising particle in the calorimeters.

%%%%%%%%%%%%%%%%%%%%%%%%%%%%%%%%%%%%%%%%%%%%%%
\begin{table}[tb]
\caption{\small 
   Summary of the relative systematic uncertainties on cross-section measurements (\%).
}
\vspace{-0.5cm}
\begin{center}
\scalebox{0.9}{%
\begin{tabular}{lccc|ccc}
\toprule
Source       & \multicolumn{3}{c|}{Forward} & \multicolumn{3}{c}{Backward} \\
             & prompt & from $b$ & inclusive & prompt & from $b$ & inclusive \\
\hline
{\it Correlated between bins} & & & & & &\\
Track reconstruction&  1.5 & 1.5 & 1.5 & 1.5 & 1.5 & 1.5 \\
Muon identification &  1.3 & 1.3 & 1.3 & 1.3 & 1.3 & 1.3 \\
Trigger      &  1.9 & 1.9 & 1.9 & 1.9 & 1.9 & 1.9 \\
Luminosity   &  1.9 & 1.9 & 1.9 & 2.1 & 2.1 & 2.1 \\
Branching fraction  &  2.2 & 2.2 & 2.2 & 2.2 & 2.2 & 2.2 \\
Track quality and radiative tail &  1.5 & 1.5 & 1.5 & 1.5 & 1.5 & 1.5 \\
Mass fit     & $3.8-6.9$ & $0.3-3.9$ & $3.2-8.2$ & $9.2-10$ & $16-20$ & $3.0-5.4$ \\
\hline
{\it Uncorrelated between bins} & & & & & &\\
Multiplicity reweighting &  0.7 & 0.7 & 0.7 & 1.7 & 1.7 & 1.7 \\
Simulation kinematics & $0.6-10$ & $0.4-10$ & $0.2-9.8$ & 1.4 & 2.4 & $0.7-23$ \\
$t_z$ fit     & $1.6-12$ & $0.3-92$ & $0.1-18$ & $1.4-7.8$ & $8.5-29$ & $0.1-17$ \\
\bottomrule
\end{tabular}%
}\end{center}
\label{tab:psi2SSystematics}
\end{table}
%%%%%%%%%%%%%%%%%%%%%%%%%%%%%%%%%%%%%%%%%%%%%%

The trigger efficiency is determined from data using a sample unbiased with respect to the trigger decision. 
The corresponding uncertainty of $1.9\%$ is taken as the systematic uncertainty due to the trigger efficiency.

To estimate the uncertainty due to reweighting the track multiplicity in simulation, the efficiency is calculated without reweighting.
The difference between cross-sections calculated with these two efficiencies
is considered as the systematic uncertainty,
which is less than $0.7\%$ in the forward sample,
and about $1.7\%$ in the backward sample.

The possible difference of the \pt and \y spectra inside each kinematic bin 
between data and simulation can introduce a systematic uncertainty.
To estimate the size of this effect the acceptance and reconstruction efficiencies have been checked by doubling the number of bins in \pt or in $y$.
The difference from the nominal binning scheme is taken as systematic uncertainty,
which is $0.2\%-10\%$ ($0.7\%-23\%$) in the forward (backward) sample.
For the backward sample the separation into prompt \psitwos and \psitwos from \bquark was not done in bins of \pt and \y due to the limited sample size.

The luminosity is determined with an uncertainty of $1.9\%$ ($2.1\%$)
for the $\pPb$ forward (backward) sample~\cite{LHCb-PAPER-2014-047}.
The uncertainty on the $\psitwos\to\mumu$ branching fraction is $2.2\%$.
The combined uncertainty related to the track quality, the vertex finding and the radiative tail is estimated to be $1.5\%$.

The uncertainty due to modelling the invariant mass distribution is estimated by using the signal shape from simulation convolved with a Gaussian function, or by replacing the exponential function by a second-order polynomial. 
The maximum differences from the nominal results are taken as the systematic uncertainties due to the mass fit.
To estimate the corresponding systematic uncertainty on the differential production cross-section due to the fixed mass resolution, the mass resolution is shifted by one standard deviation. It is found that this uncertainty is negligible.
The uncertainty due to modelling the $t_z$ distribution is estimated by fitting the signal
sample extracted from the \sPlot technique using the invariant mass alone as the discriminating variable.

%%%%%%%%%%%%%%%%%%%%%%%%%%%%%%%%%%
\section{Results}
\subsection{Cross-sections}
\begin{figure}[tb]
\begin{center}
  \includegraphics[width=0.49\textwidth]{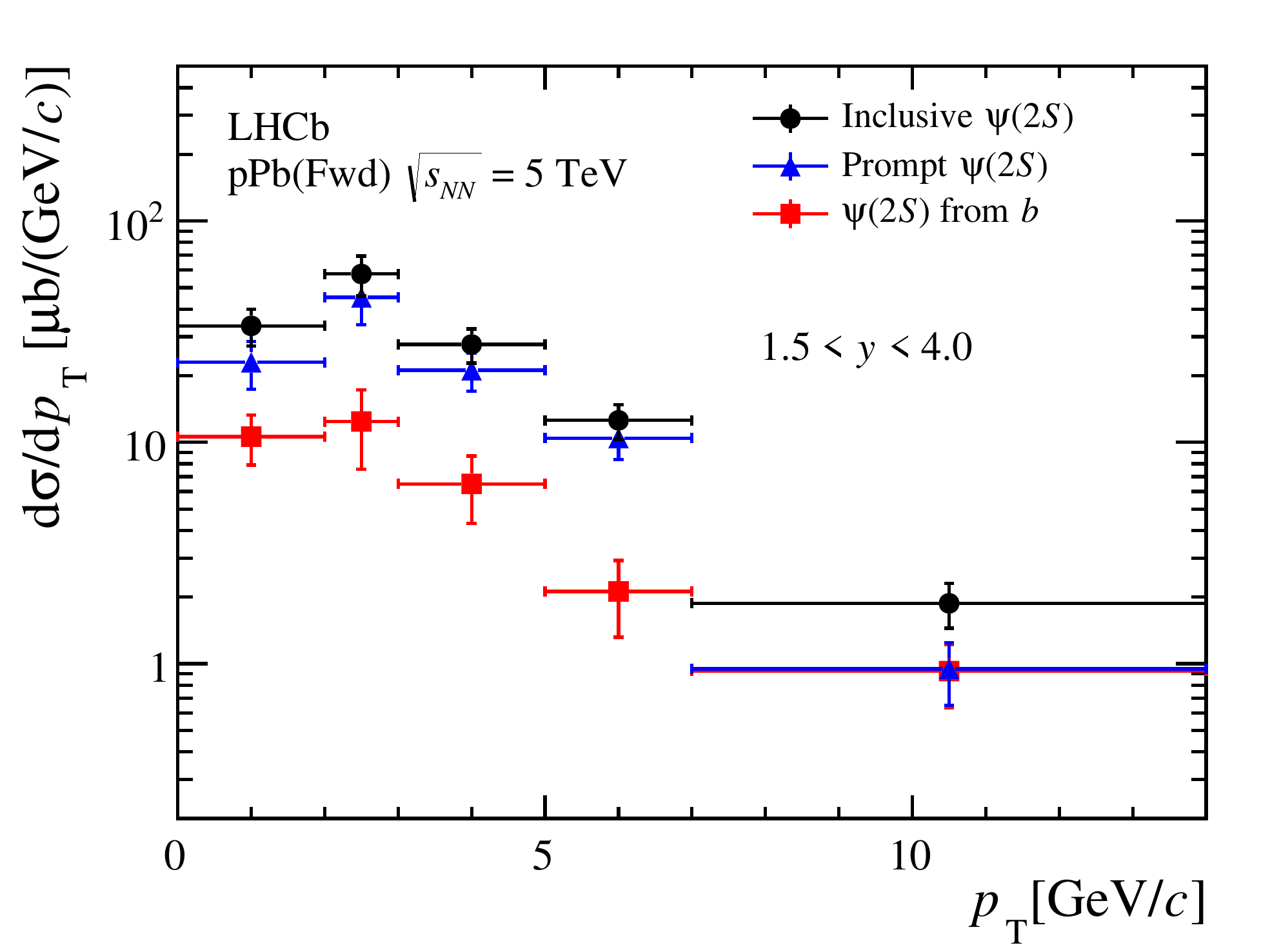}
  \includegraphics[width=0.49\textwidth]{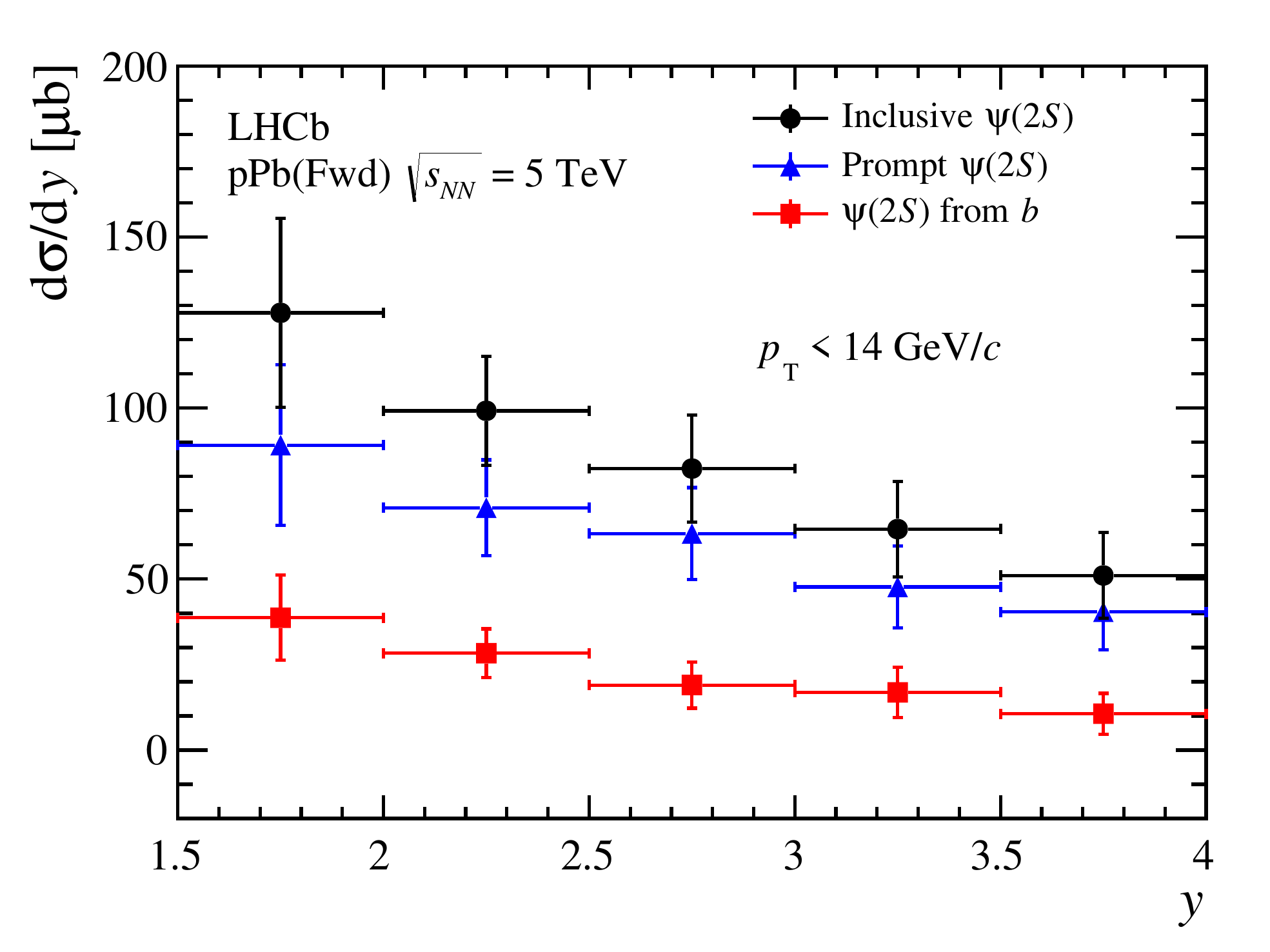}
\end{center}
\vspace*{-0.5cm}
\caption{ 
 Differential cross-section of \psitwos meson production as a function of 
 (left) \pt and (right) \y in \pPb forward collisions.
 The (black) dots represent inclusive \psitwos, the (blue) triangles indicate 
 prompt \psitwos, and the (red) squares show \psitwos from $b$.
 The error bars indicate the total uncertainties.
}
\label{fig:psi2SXsectionpAPtY_event}
\end{figure}

\begin{figure}[tb]
\begin{center}
 \includegraphics[width=0.49\linewidth]{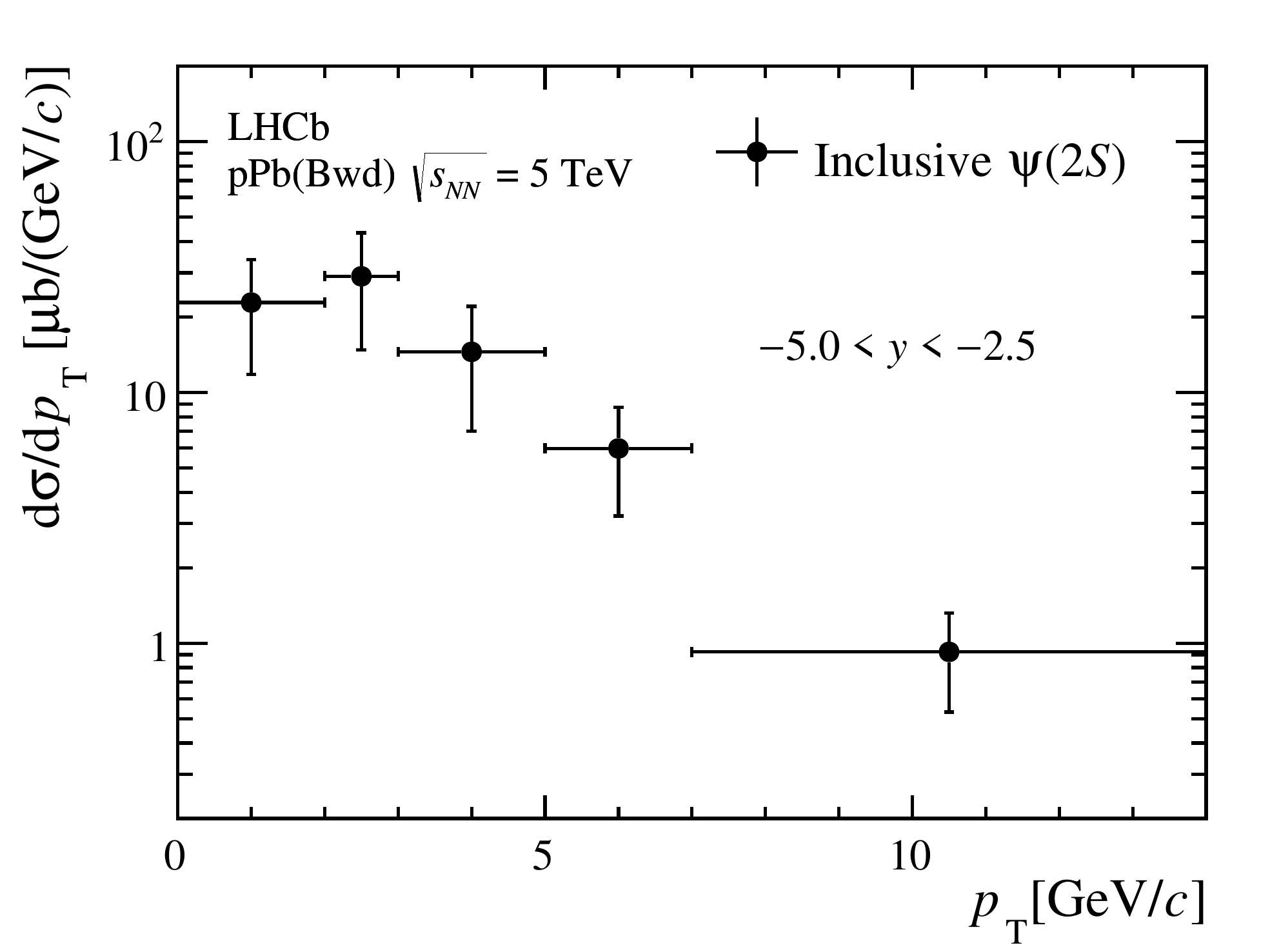}
 \includegraphics[width=0.49\linewidth]{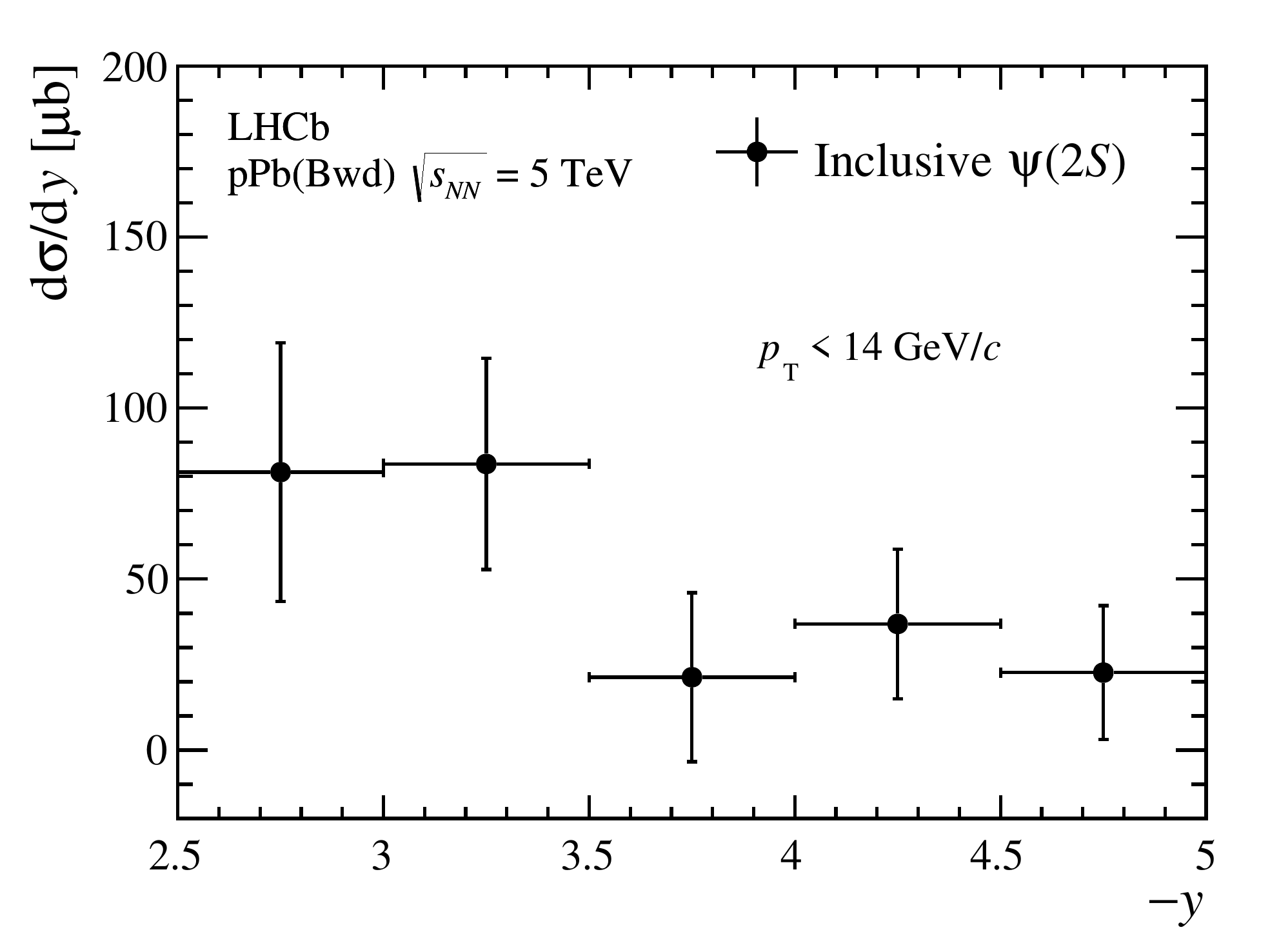}
\end{center}
\vspace*{-0.5cm}
\caption{
 Differential cross-section of \psitwos meson production as a function of (left) \pt 
 and (right) \y in \pPb backward collisions. 
 The error bars indicate the total uncertainties.
}
\label{fig:psi2SXsectionApPtY_event}
\end{figure}

The differential cross-sections of prompt \psitwos, \psitwos from $b$
and inclusive \psitwos in the \pPb forward region as functions of \pt and \y are shown 
in Fig.\,\ref{fig:psi2SXsectionpAPtY_event}. 
The differential cross-sections of inclusive \psitwos in the \pPb backward region as functions of \pt and \y are shown in Fig.\,\ref{fig:psi2SXsectionApPtY_event}.
As stated in Sect.~\ref{sec:systematics}, for the differential production cross-section in the backward data sample, no attempt is made to separate prompt \psitwos and \psitwos from \bquark due to the small statistics. 
However, these two components are separated for the integrated production cross-sections.
All these cross-sections decrease with increasing $|y|$. 

The integrated production cross-sections for prompt \psitwos, \psitwos from $b$, and their sum representing inclusive \psitwos, are given in Table~\ref{tab:psi2SIntegrated}.
To determine the forward-backward production ratio $\RFB$, the integrated production cross-sections in the common rapidity region, $2.5<|y|<4.0$, are also given in the table.
%%%%%%%%%%%%%%%%%%%%%%%%%%%%%%%%%%%%%%%%%%%
%%%%%%%%%%%%%%%%%%%%%%%%%%%%%%%%%%%%%%%%%%%%
%%%%%%%%%%%%%%%%%%%%%%%%%%%%%%%%%%%%%%%%%%%%
%%%%%%%%%%%%%%%%%%%%%%%%%% Integrated cross-sections %%%%%%%%%%%%%%
\begin{table}[b]
\caption{ 
 Integrated production cross-sections for prompt \psitwos, \psitwos from $b$, 
 and inclusive \psitwos in the forward region and the backward region. The \pt range is $\pt<14\gevc$.
 The first uncertainty is statistical and the second is systematic.
}
\vspace{-0.5cm}
\begin{center}
\scalebox{1.0}{%
\begin{tabular}{lccc}
\toprule
                 & prompt [\mymub]      & from $b$ [\mymub]  & inclusive [\mymub] \\
Forward~~~($+1.5<y<+4.0$) & $  138\pm17\pm\xx8$  & $53.7\pm7.9\pm3.6$ & $  192\pm19\pm10$ \\
Backward~($-5.0<y<-2.5$)  & $\xx93\pm25\pm10$    & $20.2\pm8.0\pm4.3$ & $  113\pm26\pm11$ \\
Forward~~~($+2.5<y<+4.0$) & $\xx65\pm10\pm\xx6$  & $21.4\pm4.5\pm1.1$ & $\xx86\pm11\pm\xx7$ \\
Backward~($-4.0<y<-2.5$)  & $\xx76\pm23\pm10$    & $13.8\pm6.9\pm5.7$ & $\xx90\pm24\pm12$ \\
\bottomrule
\end{tabular}%
}
\end{center}
\label{tab:psi2SIntegrated}
\end{table}

The production cross-sections, $\sigma(\bbbar)$, of the \bbbar pair can be obtained from 
\begin{equation}
 \sigma(\bbbar)=\sigma(\psitwos \mbox{ from } b)/2f_{b\to\psitwos}
               =\sigma(\jpsi \mbox{ from } b)/2f_{b\to\jpsi},
\end{equation}
where $f_{b\to\psitwos}$ ($f_{b\to\jpsi}$) indicates the production fraction of $b\to\psitwos X$ (\mbox{$b\to\jpsi X$}). 
The world average values are $f_{b\to\jpsi}=(1.16\pm0.10)\times10^{-2}$ and \mbox{$f_{b\to\psitwos}=(2.83\pm0.29)\times10^{-3}$}~\cite{PDG2014}.   
The production cross-sections $\sigma(\bbbar)$ obtained from the results of \jpsi and \psitwos from \bquark are shown in Table~\ref{tab:bbbar}.
The results of the $\bbbar$ cross-sections obtained
from \psitwos from \bquark are consistent with those from \jpsi from \bquark. 

In the combination of the results the partial correlation between $f_{b\to\psitwos}$ and $f_{b\to\jpsi}$ is taken into account.
The systematic uncertainties due to the muon identification, the tracking efficiency, and the track quality are considered to be fully correlated.
The systematic uncertainties due to the luminosities are partially correlated.
The averaged results are also shown in Table~\ref{tab:bbbar}.

%%%%%%%%%%%%%%%%%%%%%%%%%% sigma_bb %%%%%%%%%%%%%%
\begin{table}[tb]
\caption{ 
 Production cross-sections $\sigma(\bbbar)$ of $\bbbar$ pairs in \pPb collisions
 obtained from the production cross-sections of \jpsi and \psitwos from $b$. 
 The superscript $\psi$ denotes \jpsi or \psitwos.
 The first uncertainties are statistical,
 the second are systematic, and the third are due to the production branching fractions.
 The last row gives the averaged results, with the first uncertainty uncorrelated and the second correlated.
}
\vspace{-0.5cm}
\begin{center}
\scalebox{1.0}{%
\begin{tabular}{lcc}
\toprule
  & $\sigma_{\Fwd}(\bbbar)$ [$\mymbarn$]  & $\sigma_{\Bwd}(\bbbar)$ [$\mymbarn$] \\
  & ($\pt^\psi<14\gevc$, $1.5<y^\psi<4.0$) & ($\pt^\psi<14\gevc$, $-5.0<y^\psi<-2.5$) \\
\hline
\psitwos & $9.49\pm1.40\pm0.64\pm0.97$  & $3.57\pm1.41\pm0.76\pm0.37$ \\
\jpsi    & $7.16\pm0.18\pm0.40\pm0.62$& $5.09\pm0.29\pm0.53\pm0.44$ \\
Averaged& $7.43\pm0.56(\uncorr)\pm0.49(\corr)$ & $4.87\pm0.62(\uncorr)\pm0.32(\corr)$ \\
\bottomrule
\end{tabular}%
}
\end{center}
\label{tab:bbbar}
\end{table}

%%%%%%%%%%%%%%%%%%%%%%%%%%%%%%%%%%%%%%%%%%%%%%
\subsection{Cold nuclear matter effects}
\begin{figure}[tb]
\begin{center}
 \includegraphics[width=0.49\linewidth]{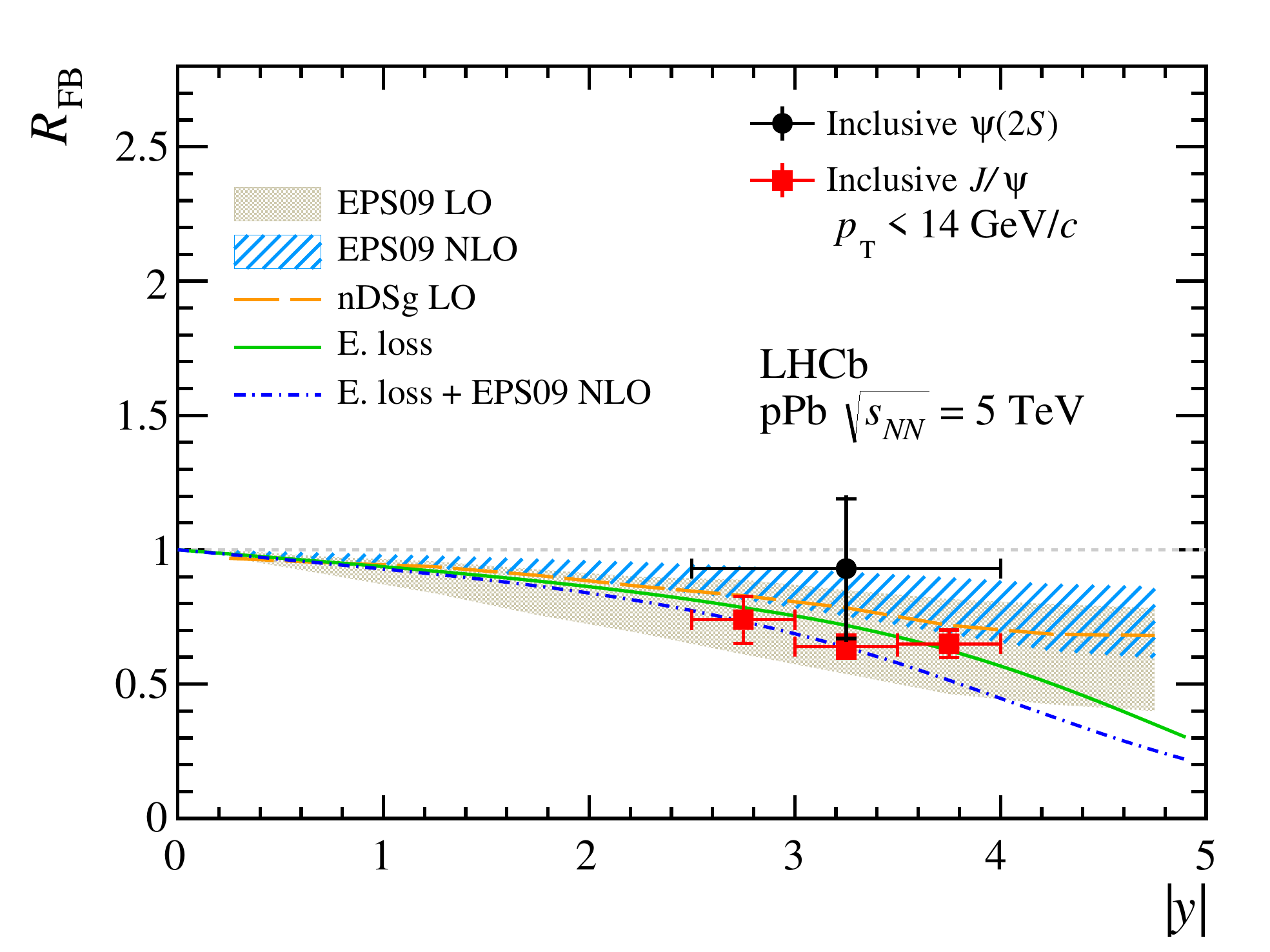}
 \includegraphics[width=0.49\textwidth]{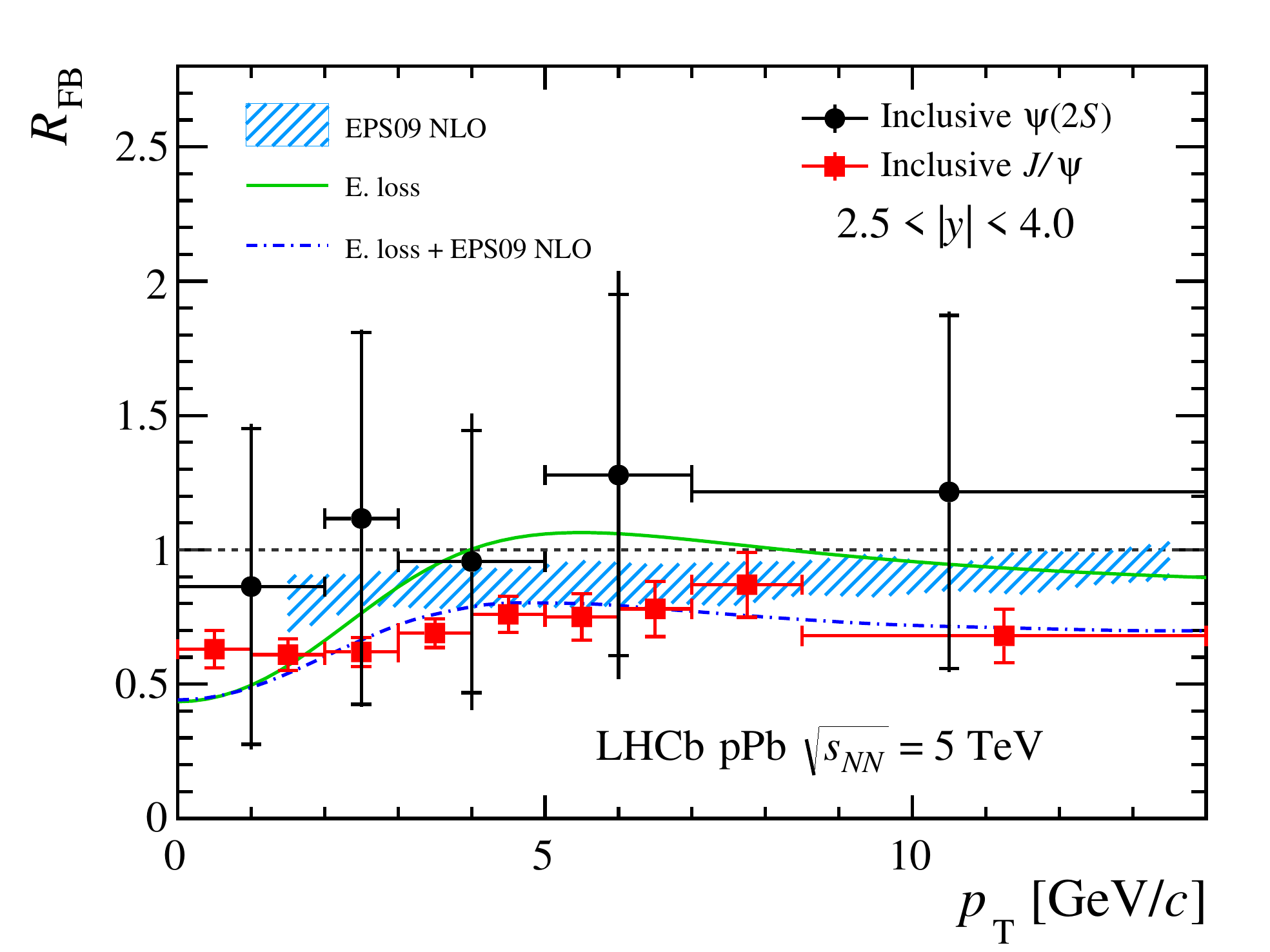}
\end{center}
\vspace*{-0.5cm}
\caption{
 Forward-backward production ratios $\RFB$ as functions of (left) $|y|$ and (right) \pt
 for inclusive $\psitwos$ mesons, 
 together with inclusive \jpsi results~\cite{LHCb-PAPER-2013-052} 
 and the theoretical predictions~\cite{Arleo:2012rs,Ferreiro:2013pua,Albacete:2013ei},
 only some of which are available for $|y|$.
 The inner error bars (delimited by the horizontal lines) 
 show the statistical uncertainties; 
 the outer ones show the statistical and systematic uncertainties added in quadrature.
}
\label{fig:RFB}
\end{figure}

Cold nuclear matter effects on \psitwos mesons can be studied with 
the production cross-sections obtained in the previous section.
As defined in Eq.~\ref{eq:RFB}, the forward-backward production ratio, $\RFB$, 
can be determined with the cross-sections in the common rapidity range ($2.5<|y|<4.0)$. 
The results are
\begin{equation*}
 \RFB(\pt<14\gevc,2.5<|y|<4.0)= 
 \begin{cases}
 0.93\pm0.29\pm0.08,\qquad \mathrm{inclusive},\\
 0.86\pm0.29\pm0.10,\qquad \mathrm{prompt},\\
 1.55\pm0.84\pm0.59,\qquad \mathrm{from}~b,
 \end{cases}
\end{equation*}
where the first uncertainties are statistical and the second systematic.
The ratios \RFB for inclusive \psitwos production as functions of \y and \pt are shown 
in Fig.\,\ref{fig:RFB}.
For comparison, the plots also show the results for 
inclusive \jpsi production~\cite{LHCb-PAPER-2013-052} 
and the theoretical predictions for \psitwos~\cite{Arleo:2012rs,Ferreiro:2013pua,Albacete:2013ei}.
The uncertainties for the theoretical predictions are obtained by taking into account minimum and maximum nuclear shadowing effects, with many of them cancelling in the ratios.
Calculations in Ref.~\cite{Ferreiro:2013pua} are based on
the Leading Order Colour Singlet Model (LO CSM)~\cite{Chang:1979nn,Baier:1981uk},
taking into account the modification effects of the gluon distribution function
in nuclei with the parameterisation EPS09~\cite{Eskola:2009uj} or nDSg~\cite{deFlorian:2003qf}.
The next-to-leading order Colour Evaporation Model
(NLO CEM)~\cite{CEM} is used in Ref.~\cite{Albacete:2013ei},
considering parton shadowing with the EPS09 parameterisation.
Reference~\cite{Arleo:2012rs} provides theoretical predictions of a coherent
parton energy loss effect both in initial and final states, with or without additional parton shadowing effects according to EPS09.
The single free parameter $q_0$ in this model is $0.055~(0.075)\gevtwofm$ when parton shadowing in the EPS09 parameterisation is (not) taken into account.
Within uncertainties the measurements agree with all these calculations.

%%%%%%%%%%%%%%%%%%% R_{FB} %%%%%%%%%%%%%
%%%%%%%%%%%%%%%%%%%%%%%%%%%%%%%%%%%%%%%%%%

\begin{figure}[tb]
\begin{center}
 \includegraphics[width=0.49\textwidth]{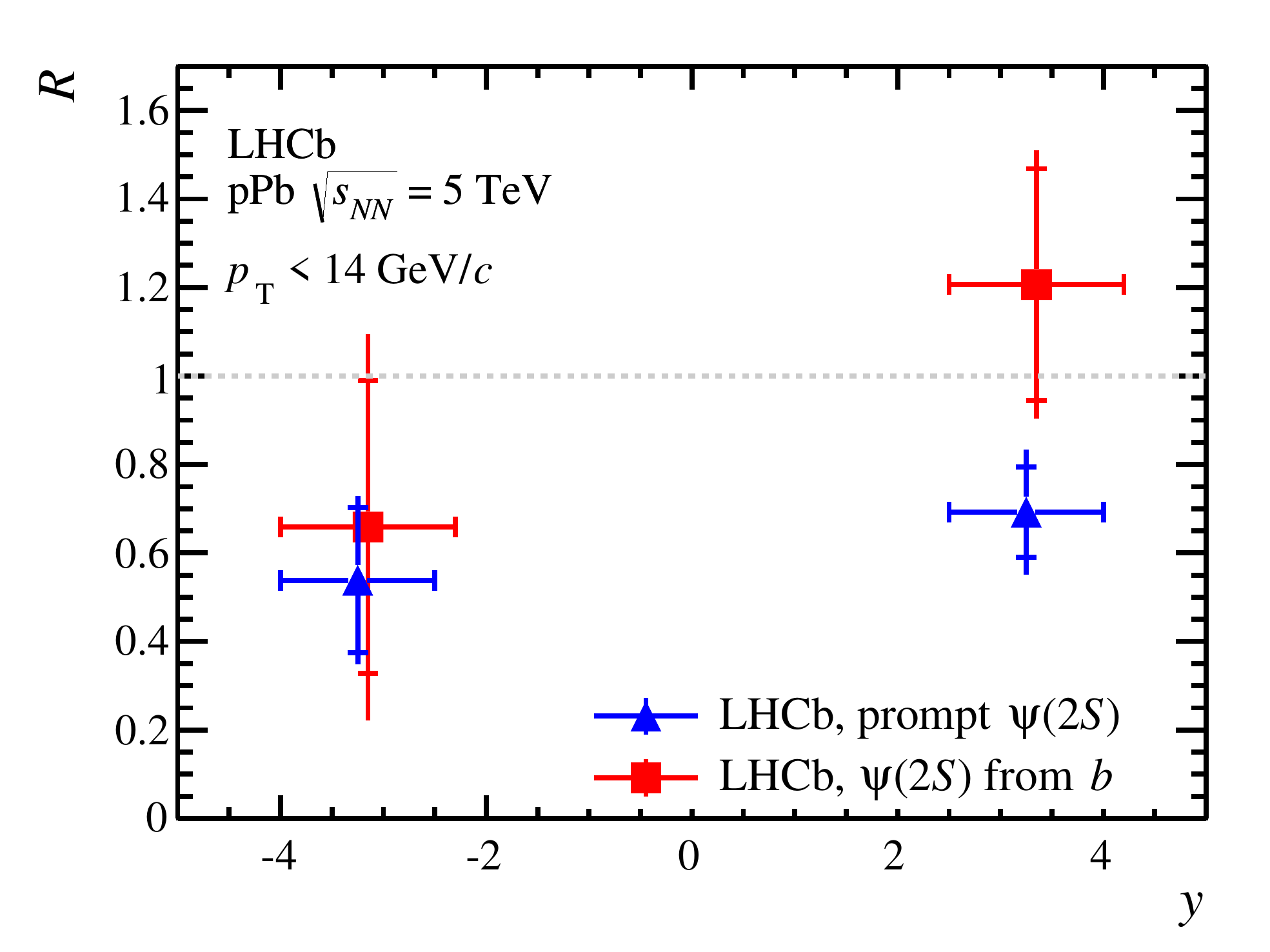}
 \includegraphics[width=0.49\textwidth]{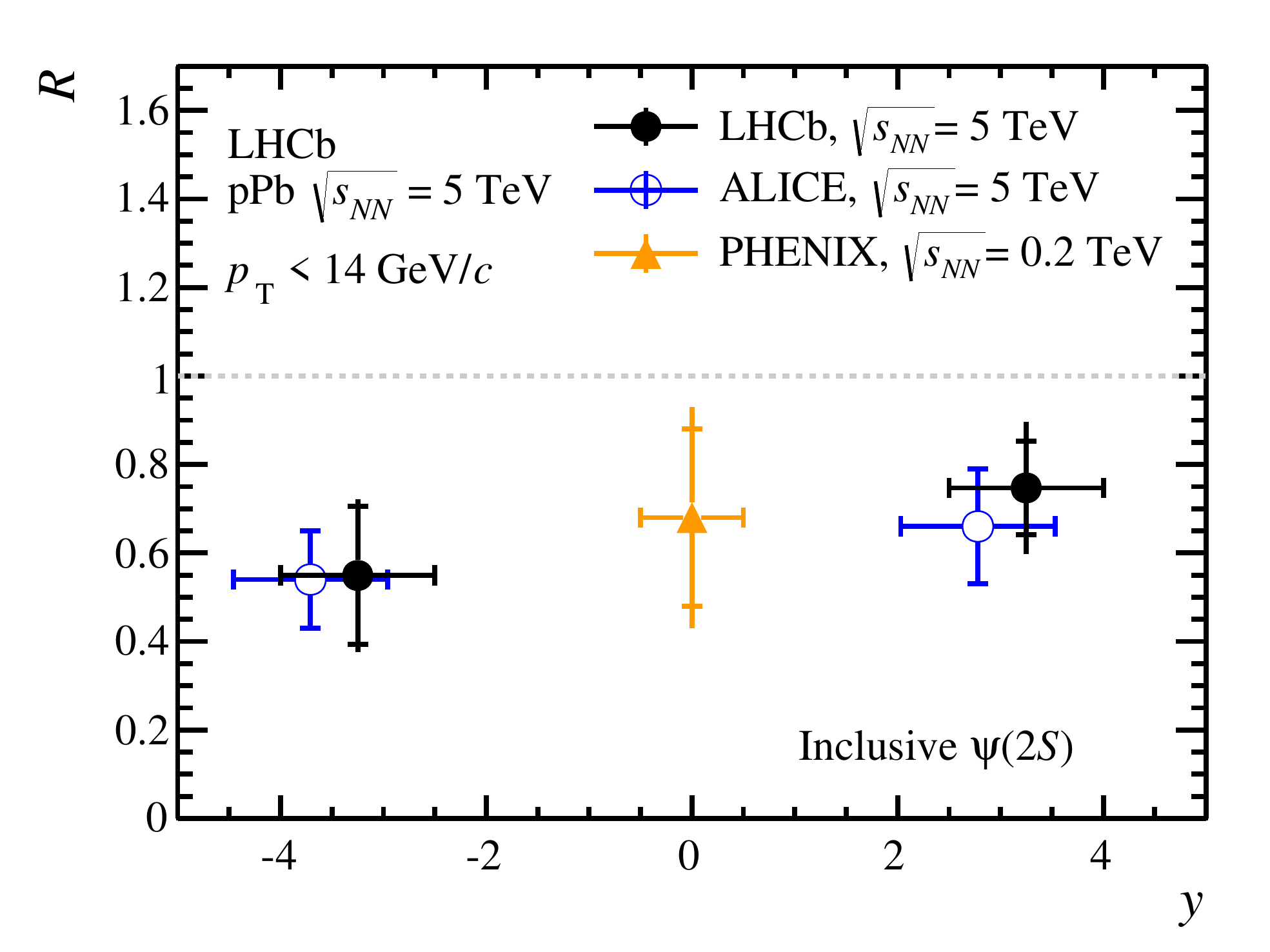}
\end{center}
\vspace*{-0.5cm}
\caption{
Ratio (left) between nuclear modification factors of \psitwos and \jpsi as a function of $y$ for prompt $\psitwos$ mesons and \psitwos from \bquark. The blue triangles represent prompt \psitwos and the red rectangles indicate \psitwos from \bquark.
Ratio (right) between nuclear modification factors of \psitwos and \jpsi as a function of $y$ for inclusive $\psitwos$ mesons. 
 The black dots show the LHCb result,
 the hollow circles indicate the ALICE result,
 and the brown triangle is the PHENIX result at $0.2\protect\tev$.
 The inner error bars (delimited by the horizontal lines) 
 show the statistical uncertainties; 
 the outer ones show the statistical and systematic uncertainties added in quadrature.
}
\label{fig:R}
\end{figure}

To obtain the nuclear modification factor $\RpPb$, 
the \psitwos production cross-section in $pp$ collisions at $5\tev$ is needed, 
which is not yet available.
However, it is reasonable to assume that
\begin{equation}
 \label{eq:JpsiTwosratio}
 \frac{\sigma_{\pp}^{\jpsi}(5\tev)}{\sigma_{\pp}^{\psitwos}(5\tev)}
 =\frac{\sigma_{\pp}^{\jpsi}(7\tev)}{\sigma_{\pp}^{\psitwos}(7\tev)},
\end{equation}
where $\sigma_{pp}$ indicates the production cross-section of \jpsi or \psitwos in \pp collisions.
The systematic uncertainty due to this assumption is taken to be negligible compared with the statistical uncertainties in this analysis.
The ratio $R$ of nuclear matter effects between \psitwos and \jpsi can then be determined as
\begin{equation}
\label{eq:RelativeSuppression}
 R\equiv\frac{\RpPb^{\psitwos}}{\RpPb^{\jpsi}}
=\frac{\sigma_{\pPb}^{\psitwos}(5\tev)}{\sigma_{\pPb}^{\jpsi}(5\tev)}
 \times\frac{\sigma_{\pp}^{\jpsi}(5\tev)}{\sigma_{\pp}^{\psitwos}(5\tev)}
=\frac{\sigma_{\pPb}^{\psitwos}(5\tev)}{\sigma_{\pPb}^{\jpsi}(5\tev)}
 \times\frac{\sigma_{\pp}^{\jpsi}(7\tev)}{\sigma_{\pp}^{\psitwos}(7\tev)},
\end{equation}
where $\RpPb^{\psitwos}$ and $\RpPb^{\jpsi}$ are the nuclear modification factors for \psitwos and \jpsi.
The ratio $R$ indicates whether there is relative suppression between \psitwos and \jpsi 
production in the collisions. If $R$ is less than unity, it suggests that 
the suppression of \psitwos mesons due to nuclear matter effects in \pPb collisions
is stronger than that of \jpsi mesons.
Using previous LHCb measurements~\cite{LHCb-PAPER-2011-003,LHCb-PAPER-2011-045,LHCb-PAPER-2013-052}, the values of $R$ for prompt \psitwos, \psitwos from \bquark and inclusive \psitwos are calculated. The results are shown in Fig.\,\ref{fig:R}, together with those
from ALICE~\cite{Abelev:2014zpa} and PHENIX~\cite{Adare:2013ezl}.
The LHCb measurement is consistent with ALICE, which is in a comparable kinematic range.
All results suggest a stronger suppression for prompt \psitwos mesons than that for prompt \jpsi mesons.

\begin{figure}[tb]
\begin{center}
 \includegraphics[width=0.49\textwidth]{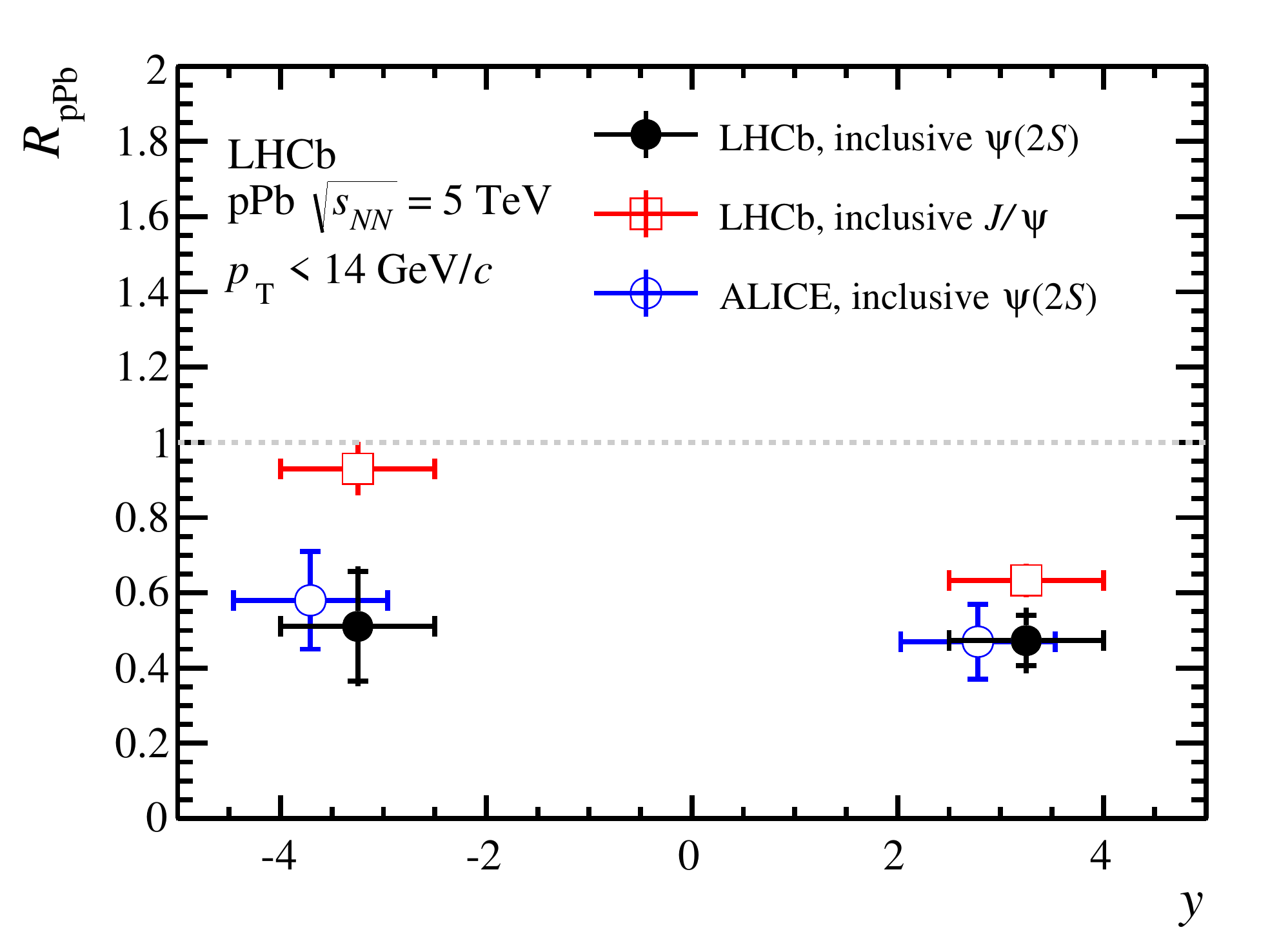}
\end{center}
\vspace*{-0.5cm}
\caption{
 Nuclear modification factor \RpPb as a function of $y$ for inclusive $\psitwos$ and \jpsi mesons. 
 The black dots represent the \psitwos result,
 the red squares indicate the \jpsi result,
 and the blue hollow circles show the ALICE result for \psitwos.
 The inner error bars (delimited by the horizontal lines) 
 show the statistical uncertainties; 
 the outer ones show the statistical and systematic uncertainties added in quadrature.
}
\label{fig:RpPb_inclusive}
\end{figure}

\begin{figure}[tb]
\begin{center}
 \includegraphics[width=0.49\textwidth]{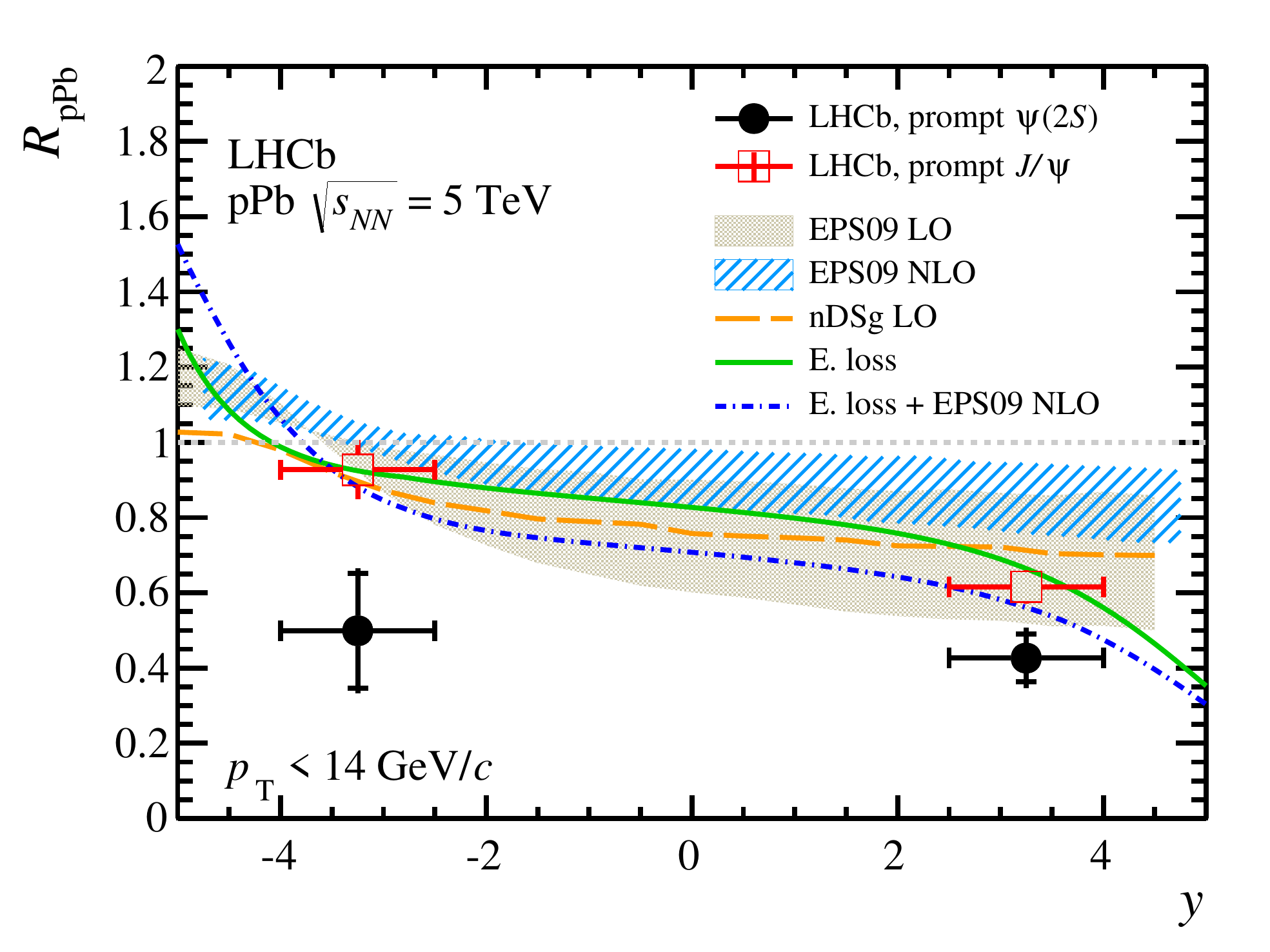}
 \includegraphics[width=0.49\textwidth]{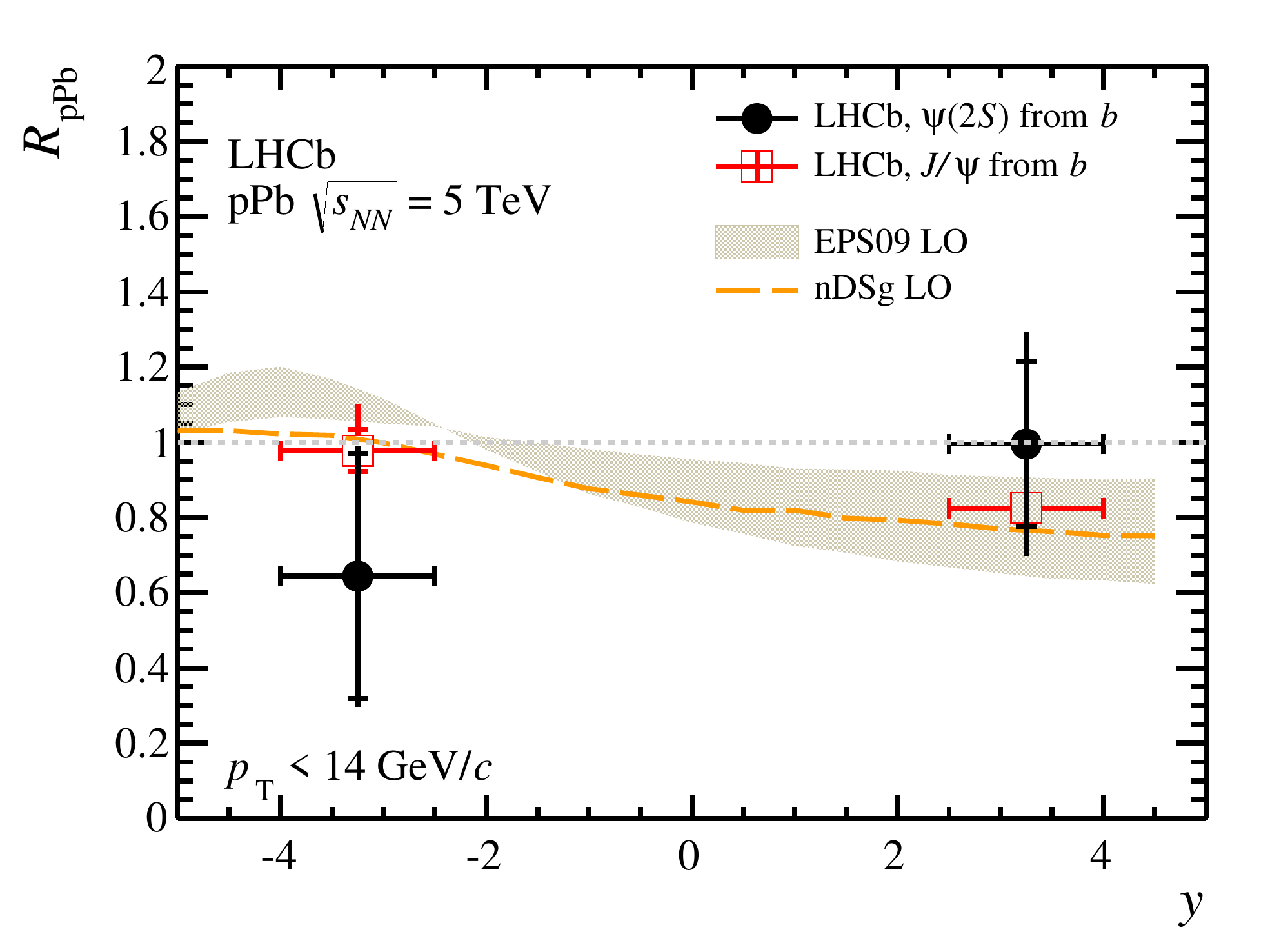}
\end{center}
\vspace*{-0.5cm}
\caption{
 Nuclear modification factor \RpPb as a function of $y$ for (left) prompt $\psitwos$ 
 and (right) \psitwos from $b$, together with the theoretical predictions from 
 (yellow dashed line and brown band) Refs.~\cite{Ferreiro:2013pua,delValle:2014wha},
 (blue band) Ref.~\cite{Albacete:2013ei}, and
 (green solid and blue dash-dotted lines) Ref.~\cite{Arleo:2012rs},
 where only those from Ref.~\cite{delValle:2014wha} are available for \psitwos from \bquark.
 The inner error bars (delimited by the horizontal lines) 
 show the statistical uncertainties; 
 the outer ones show the statistical and systematic uncertainties added in quadrature.
}
\label{fig:RpPb_prompt_b}
\end{figure}

The nuclear modification factor of \psitwos, $\RpPb^{\psitwos}$,
can be expressed in terms of $\RpPb^{\jpsi}$ and $R$
\begin{equation}
\label{eq:RpPb_Jpsi_twos}
 \RpPb^{\psitwos}=\RpPb^{\jpsi}\times{}R\,.
\end{equation}
The nuclear modification factor $\RpPb^{\jpsi}$ was determined in a previous measurement~\cite{LHCb-PAPER-2013-052}.
The result for inclusive \psitwos is shown in Fig.\,\ref{fig:RpPb_inclusive}.
For comparison, the inclusive \jpsi result from previous measurements~\cite{LHCb-PAPER-2013-052}
and the result from ALICE~\cite{Abelev:2014zpa} are also shown in the plot. The LHCb measurement is consistent with ALICE.
The results for prompt \psitwos and \psitwos from $b$ are shown in Fig.\,\ref{fig:RpPb_prompt_b},
suggesting that in \pPb collisions the suppression of prompt \psitwos mesons is stronger than that of prompt \jpsi mesons. For \psitwos from $b$, no conclusion can be made because of the limited sample size.
Figure~\ref{fig:RpPb_prompt_b} also shows 
several theoretical predictions~\cite{Arleo:2012rs,Ferreiro:2013pua,Albacete:2013ei,delValle:2014wha},
where only those from Ref.~\cite{delValle:2014wha} are available for \psitwos from \bquark.
For prompt \psitwos, stronger suppression is seen in the data than expected by the theoretical calculations.
Final-state effects, such as the interaction of the \ccbar pair with the dense medium created in the collisions, could be involved~\cite{Ferreiro:2014bia}.

%%%%%%%%%%%%%%%%%%% R_{FB} %%%%%%%%%%%%%
%%%%%%%%%%%%%%%%%%%%%%%%%%%%%%%%%%%%%%%%%%
%%%%%%%%%%%%%%%%%%% R_{pPb} inclusive  %%%%%%%%%%%%%
%%%%%%%%%%%%%%%%%%%%%%%%%%%%%%%%%%%%%%%%%%
%%%%%%%%%%%%%%%%%%% R_{pPb} %%%%%%%%%%%%%
%%%%%%%%%%%%%%%%%%%%%%%%%%%%%%%%%%%%%%%%%%

%%%%%%%%%%%%%%%%%%%%%%%%%

%%%%%%%%%%%%%%%%
\section{Conclusions}
\label{sec:Conclusions}
The production cross-sections of prompt $\psitwos$ mesons and those from $b$-hadron 
decays are studied in $\pPb$ collisions
with the \lhcb detector. The nucleon-nucleon centre-of-mass energy
in the collisions is $\sqrt{\sNN}=5\tev$.
The measurement is performed as a function of the transverse momentum
and rapidity of $\psitwos$ mesons
in the region $\pt<14\gevc$ and $1.5<y<4.0$ (forward) and $-5.0<y<-2.5$ (backward).
The $\bbbar$ production cross-sections in \pPb collisions are extracted using the results of \psitwos from $b$ and \jpsi from $b$.
The forward-backward production ratio \RFB is determined separately for prompt \psitwos mesons and those from \bquark-hadron decays. 
These results show agreement within uncertainties with available theoretical predictions. 
The nuclear modification factor \RpPb is also determined separately for prompt \psitwos mesons and \psitwos from \bquark. 
These results show that prompt \psitwos mesons are significantly more suppressed than prompt \jpsi mesons in the backward region; the results are not well described by theoretical predictions based on shadowing and energy loss mechanisms.

\clearpage

\section*{Acknowledgements}
\noindent We express our gratitude to our colleagues in the CERN
accelerator departments for the excellent performance of the LHC. We
thank the technical and administrative staff at the LHCb
institutes. We acknowledge support from CERN and from the national
agencies: CAPES, CNPq, FAPERJ and FINEP (Brazil); NSFC (China);
CNRS/IN2P3 (France); BMBF, DFG and MPG (Germany); INFN (Italy); 
FOM and NWO (The Netherlands); MNiSW and NCN (Poland); MEN/IFA (Romania); 
MinES and FANO (Russia); MinECo (Spain); SNSF and SER (Switzerland); 
NASU (Ukraine); STFC (United Kingdom); NSF (USA).
We acknowledge the computing resources that are provided by CERN, IN2P3 (France), KIT and DESY (Germany), INFN (Italy), SURF (The Netherlands), PIC (Spain), GridPP (United Kingdom), RRCKI and Yandex LLC (Russia), CSCS (Switzerland), IFIN-HH (Romania), CBPF (Brazil), PL-GRID (Poland) and OSC (USA). We are indebted to the communities behind the multiple open 
source software packages on which we depend.
Individual groups or members have received support from AvH Foundation (Germany),
EPLANET, Marie Sk\l{}odowska-Curie Actions and ERC (European Union), 
Conseil G\'{e}n\'{e}ral de Haute-Savoie, Labex ENIGMASS and OCEVU, 
R\'{e}gion Auvergne (France), RFBR and Yandex LLC (Russia), GVA, XuntaGal and GENCAT (Spain), The Royal Society, Royal Commission for the Exhibition of 1851 and the Leverhulme Trust (United Kingdom).

\addcontentsline{toc}{section}{References}
\setboolean{inbibliography}{true}
\bibliographystyle{LHCb}
\bibliography{main,LHCb-PAPER,LHCb-CONF,LHCb-DP,LHCb-TDR,MyBib}

% Author List ---------------------------
 
\newpage
%%%%%%%%%%%%%%%%%%%%%%%%%%%%%%%%%%%%%%%%%%
\centerline{\large\bf LHCb collaboration}
\begin{flushleft}
\small
R.~Aaij$^{39}$, 
C.~Abell\'{a}n~Beteta$^{41}$, 
B.~Adeva$^{38}$, 
M.~Adinolfi$^{47}$, 
A.~Affolder$^{53}$, 
Z.~Ajaltouni$^{5}$, 
S.~Akar$^{6}$, 
J.~Albrecht$^{10}$, 
F.~Alessio$^{39}$, 
M.~Alexander$^{52}$
% ...

{\footnotesize \it
% ...
$ ^{5}$Clermont Universit\'{e}, Universit\'{e} Blaise Pascal, CNRS/IN2P3, LPC, Clermont-Ferrand, France\\
$ ^{6}$CPPM, Aix-Marseille Universit\'{e}, CNRS/IN2P3, Marseille, France\\
% ...
$ ^{10}$Fakult\"{a}t Physik, Technische Universit\"{a}t Dortmund, Dortmund, Germany\\
% ...
$ ^{38}$Universidad de Santiago de Compostela, Santiago de Compostela, Spain\\
$ ^{39}$European Organization for Nuclear Research (CERN), Geneva, Switzerland\\
% ...
$ ^{41}$Physik-Institut, Universit\"{a}t Z\"{u}rich, Z\"{u}rich, Switzerland\\
% ...
$ ^{47}$H.H. Wills Physics Laboratory, University of Bristol, Bristol, United Kingdom\\
% ...
$ ^{52}$School of Physics and Astronomy, University of Glasgow, Glasgow, United Kingdom\\
$ ^{53}$Oliver Lodge Laboratory, University of Liverpool, Liverpool, United Kingdom\\
% ...
}
\end{flushleft}
%%%%%%%%%%%%%%%%%%%%%%%%%%%%%%%%%%%%%%%%%%

\end{document}